\documentclass{aastex}
\usepackage{emulateapj5}

\begin{document}

\shortauthors{Gordon et al.\ 2002}
\shorttitle{Comparison of SMC, LMC, \& MW Extinction Curves}

\slugcomment{}

\title{A Quantitative Comparison of SMC, LMC, and Milky Way 
UV to NIR Extinction Curves\altaffilmark{1}} 

\author{Karl D.\ Gordon\altaffilmark{2},
   Geoffrey C.\ Clayton\altaffilmark{3},
   K.\ A.\ Misselt\altaffilmark{2},
   Arlo U.\ Landolt\altaffilmark{3},
   \& Michael J.\ Wolff\altaffilmark{4}}

\altaffiltext{1}{Partially based on observations made with the
   NASA/ESA Hubble Space Telescope, obtained at the Space Telescope
   Science Institute, which is operated by the Association of
   Universities for Research in Astronomy, Inc., under NASA contract
   NAS 5-26555. These observations are associated with proposal
   \#8198}
\altaffiltext{2}{Steward Observatory, University of Arizona,
   Tucson, AZ 85721; (kgordon,kmisselt)@as.arizona.edu}
\altaffiltext{3}{Department of Physics \& Astronomy, Louisiana State
   University, Baton Rouge, LA 70803; gclayton@fenway.phys.lsu.edu,
   landolt@rouge.phys.lsu.edu}
\altaffiltext{4}{Space Science Institute, 1540 30th Street, Suite 23
   Boulder, CO 80303-1012; wolff@colorado.edu}

\begin{abstract} 
We present an exhaustive, quantitative comparison of all of the known
extinction curves in the Small and Large Magellanic Clouds (LMC and
SMC) with our understanding of the general behavior of Milky Way
extinction curves.  The $R_V$ dependent CCM relationship and the
sample of extinction curves used to derive this relationship is used
to describe the general behavior of Milky Way extinction
curves.  The ultraviolet portion of the SMC and LMC extinction curves
are derived from archival IUE data, except for one new SMC extinction
curve which was measured using HST/STIS observations.  The optical
extinction curves are derived from new (for the SMC) and literature
UBVRI photometry (for the LMC).  The near-infrared extinction curves
are calculated mainly from 2MASS photometry supplemented with DENIS
and new JHK photometry.  For each extinction curve, we give $R_V =
A(V)/E(B-V)$ and $N(HI)$ values which probe the same dust column as
the extinction curve.  We compare the properties of the SMC and LMC
extinction curves with the CCM relationship three different ways: each
curve by itself, the behavior of extinction at different wavelengths
with $R_V$, and behavior of the extinction curve FM fit parameters
with $R_V$.  As has been found previously, we find that a small number
of LMC extinction curves are consistent with the CCM relationship, but
majority of the LMC and all of the SMC curves do not follow the CCM
relationship.  For the first time, we find that the CCM relationship
seems to form a bound on the properties of all of the LMC and SMC
extinction curves.  This result strengthens the picture of dust
extinction curves exhibit a continuum of properties between those
found in the Milky Way and the SMC Bar.  Tentative evidence based on
the behavior of the extinction curves with dust-to-gas ratio suggests
that the continuum of dust extinction curves is possibly caused by the
environmental stresses of nearby star formation activity.
\end{abstract}


\section{Introduction \label{sec_intro}}

One of the main tools used in the study of dust grain properties is
extinction curves, particularly ultraviolet (UV) extinction curves.
The UV is where dust extinction is strongest and shows the large
variations from region to region in the Milky Way
\citep{wit84,aie88,fit90,cla00}, Large Magellanic Cloud (LMC)
\citep{cla85,fit86,mis99}, and Small Magellanic Cloud (SMC)
\citep{leq82,pre84,gor98}.  

The work of \citet{car89} found that most of the variation in Milky
Way extinction curves could be described by an empirically relationship
based on the single parameter $R_V = A_V/E(B-V)$.  This was a major
step forward in our understanding of dust properties and was possible
only due to the existence in the literature of near-infrared
photometry for a subset of stars in the \citet{fit90} sample.  This
allowed \citet{car89} to determine $R_V$ values and transform the
\citet{fit90} extinction curves to an absolute scale (i.e., normalized
to $A_V$ instead of $E(B-V)$).  These $A_V$ normalized curves had
variations which were correlated with $R_V$ allowing \citet{car89} to
empirically derive the $R_V$ dependent CCM relationship.  Since the
$R_V$ value is a rough measure of average dust grain size, this 
gave a physical basis for the variations in extinction curves.  It is
worth noting that significant, small deviations from the CCM
relationship are seen for individual sightlines \citep{mat92}.

The one caveat on the result of \citet{car89} is that the
strength of UV dust extinction limits the measurement of UV dust
extinction curves to low to moderate reddening sightlines.  This
results in a significant bias in measured UV extinction curves in the
Milky Way to regions surrounding the Sun.  In the one of the largest,
detailed study of UV dust extinction curves in the Milky Way to date,
\citet{fit90} presented curves for 78 sightlines and the average
distance probed was 1.3 kpc.  The CCM relationship was derived from a
subset of the \citet{fit90} sample and, as such, this relationship
might only be valid for dust in our region of the Milky Way.
\citet{cla00} tested the validity of the CCM relationship for a larger
region of the Milky Way by measuring UV extinction curves along very
low-density sightlines.  The 26 extinction curves in their sample had
an average distance of 5.2 kpc.  They found that 19 of the 26
extinction curves in their sample were qualitatively consistent with
the CCM relationship.  The remaining 7 curves had shapes which are not
described by the CCM relationship and were qualitatively similar to
those seen in the part of the Large Magellanic Cloud (LMC)
\citep{mis99} associated with the LMC2 supershell (near the 30 Dor
star formation region).  These 7 curves were all clustered in the same
region in the sky and this sightline through the galaxy displays
evidence for shocked dust \citep{cla00}.

The CCM relationship seems to be a good description of Milky Way
extinction curves with a few exceptions.  This raises the question:
Does the CCM relationship describe the dust outside of the Milky Way?
In other words, are the extinction curves in other galaxies
quantitatively similar to those in the Milky Way?  Only in the
Magellanic Clouds can this question be answered as these are the only
two galaxies with reliably measured UV extinction curves.  A full
answer this question requires $R_V$ values for all the Magellanic
Cloud extinction curves and this has only been possible with the
recent release of the 2MASS observations of the Magellanic Clouds.
This is the motivation for this paper.  Even without all the needed
$R_V$ measurements, previous work has gone a long way in answering the
above question.  It was quickly realized with the first few measured
extinction curves in the LMC and SMC the that Clouds had curves which
were similar to Milky Way curves as well as curves which were quite
different.  For example, the sightlines towards the LMC star
Sk~-69~108 \citep{nan78} and the SMC star AzV~456 \citep{leq82}
display Milky Way-like extinction curves.  On the other hand,
sightlines near 30-Dor in the LMC \citep{cla85,fit86} and in the star
forming Bar of the SMC \citep{pre84} show definite non-Milky Way-like
extinction curves especially in their 2175~\AA\ bump and far-UV rise
strengths.  

In order to move from a qualitative to a quantitative comparison of
Magellanic Cloud and Milky Way extinction curves, $R_V$ values are
needed for each extinction curve.  This allows for the normalization
of the extinction curves by $A_V$ instead of the usual $E(B-V)$ which
the \citet{car89} work proved was vitally important in understanding
the true differences between extinction curves.  While, the studies of
\citet{gor98} and \citet{mis99} concentrated on deriving all the UV
extinction curves possible with {\em International Ultraviolet
Explorer} (IUE) archival data in the SMC and LMC, respectively, they
also presented $R_V$ values for a subset of the SMC and LMC curves.
Ideally, the $R_V$ values for each extinction curve should be derived
from near-infrared photometry of both the reddened and comparison
stars which make up each curve.  This ensures that the measured
extinction curve and $R_V$ value correspond to the same dust column
and both are similarly corrected for foreground Milky Way dust.  Due
to the paucity of near-infrared photometry for their reddened and
(especially) comparison stars, a majority of the $R_V$ values
presented in \citet{gor98} and \citet{mis99} were based on assumed
near-infrared intrinsic colors.  In addition, the reddened stars'
colors were not corrected for Milky Way foreground dust.  This
interjected a significant error in the $R_V$ values.  For example, the
foreground extinction can be up to 25\% of the total extinction for
the LMC extinction curves \citep{mis99}.  This results in significant
differences between $R_V$ values reported in these two studies and
previous work \citep{mor82}.  The release of the 2MASS data for
the Magellanic Clouds makes it possible to correctly and accurately
compute $R_V$ values for all the known extinction curves in the
Magellanic Clouds \citep{gor98,mis99} using the 2MASS near-infrared
photometry of the reddened and comparison stars.

We combine archival International Ultraviolet Explorer (IUE)
ultraviolet spectra, optical photometry, and the 2MASS \citep{skr97}
and DENIS \citep{epc99, cio00} near-infrared photometry with new
Hubble Space Telescope (HST) Space Telescope Imaging Spectrograph
(STIS) ultraviolet spectroscopy and optical photometry to derive
ultraviolet through near-infrared extinction curves for 24 sightlines
in the Magellanic Clouds in \S\ref{sec_data}.  In the same section, we
also measure the $R_V$ and \ion{H}{1} column for all 24 extinction
curves.  In \S\ref{sec_discuss}, we quantitatively compare Milky Way
and Magellanic Cloud extinction curves as well as discuss various
average Magellanic Cloud extinction curves and the existence or lack
thereof of the 2175~\AA\ bump in the SMC Bar.

\section{Data \label{sec_data}}

We present data on all the sightlines in the Magellanic Clouds which
have UV extinction curves.  The sightlines include 23 based on IUE
data which have been published previously \citep{gor98, mis99} and one
based on STIS data which are published for the first time in this
paper.  For each sightline, we have gathered UV spectra and optical
and near-infrared photometry for both the reddened and comparison
stars.  From these data, we have constructed UV to near-IR extinction
curves using the standard pair method and measured $R_V$ and $N(HI)$
values for each sightline.

\subsection{Optical and Near-Infrared Photometry}

The optical and near-infrared photometry for the reddened and
comparison stars is given in Table~\ref{tab_star_data}.  UBVRI
photometric data for all of the SMC stars as well as two of the LMC
stars (Sk -67 36 and Sk -68 26), were obtained on observing runs
during 1998 August and September, 1999 January, and 2001 August.  The
data were acquired at the 1.5-m telescope of the Cerro Tololo
Inter-American Observatory (CTIO).  A C31034A GaAs photomultiplier,
UBVRI filter set \#3, and the standard photoelectric data acquisition
system were used.  Extinction and transformation relations, including
non-linear transformation relations were applied to the instrumental
data.  The final magnitudes and color indices are on the photometric
system defined by \citet{lan92}.  The UBV optical photometry for the
remainder of the LMC stars was taken from \citet{mis99}.

The JHK near-infrared photometry for most of the SMC and LMC stars was
taken from the results of the 2MASS project \citep{skr97}.  The JHK
photometry for the SMC stars AzV~456 and AzV~462 was taken from
\citet{bou85}.  For two of the three LMC stars without 2MASS
photometry, we used the DENIS \citep{epc99, cio00} JK photometry
converted to the 2MASS system \citep{cut00}.  For the one LMC star
without 2MASS or DENIS photometry due to a nearby saturating star, we
used JHK images of Sk~-68~26 and its surrounding region which were
taken using the Cerro Tololo InfraRed IMager (CIRIM) at the CTIO 1.5m.
These images were taken on 15 Dec 1999 in non-photometric weather and
each band was observed 4 times offsetting between positions for a
total exposure time of 4 seconds.  The field-of-view of the coadded
images was approximately $200\arcsec \times 200\arcsec$.  In this
region, 5 of the bright stars have 2MASS observations and we used
these stars to calibrate the JHK fluxes for Sk~-68~26 using
differential photometry.  The uncertainty in these measurements was
calculated as the standard deviation of the mean of the the 5
measurements of the JHK magnitudes of Sk~-68~26, one measurement per
2MASS star.

\subsection{Ultraviolet Spectra}

The UV spectra for all but two of the stars in this paper were taken
from archival IUE observations.  The specific IUE observations we used
are given by \citet{gor98} for the SMC stars and \citet{mis99} for the
LMC stars.  The individual IUE observations were downloaded from the
MAST archive at Space Telescope Science Institute (STScI) and were
coadded to produce a single spectrum from 1150 to 3225~\AA\ with a
resolution of approximately 400.  The calibration of these spectra was
improved using the results of \citet{mas00} who found that the
signal-to-noise of IUE low-dispersion data could be significantly
improved over that provided in the IUE archive.

\begin{figure*}[tbp]
\epsscale{2.1}
\plottwo{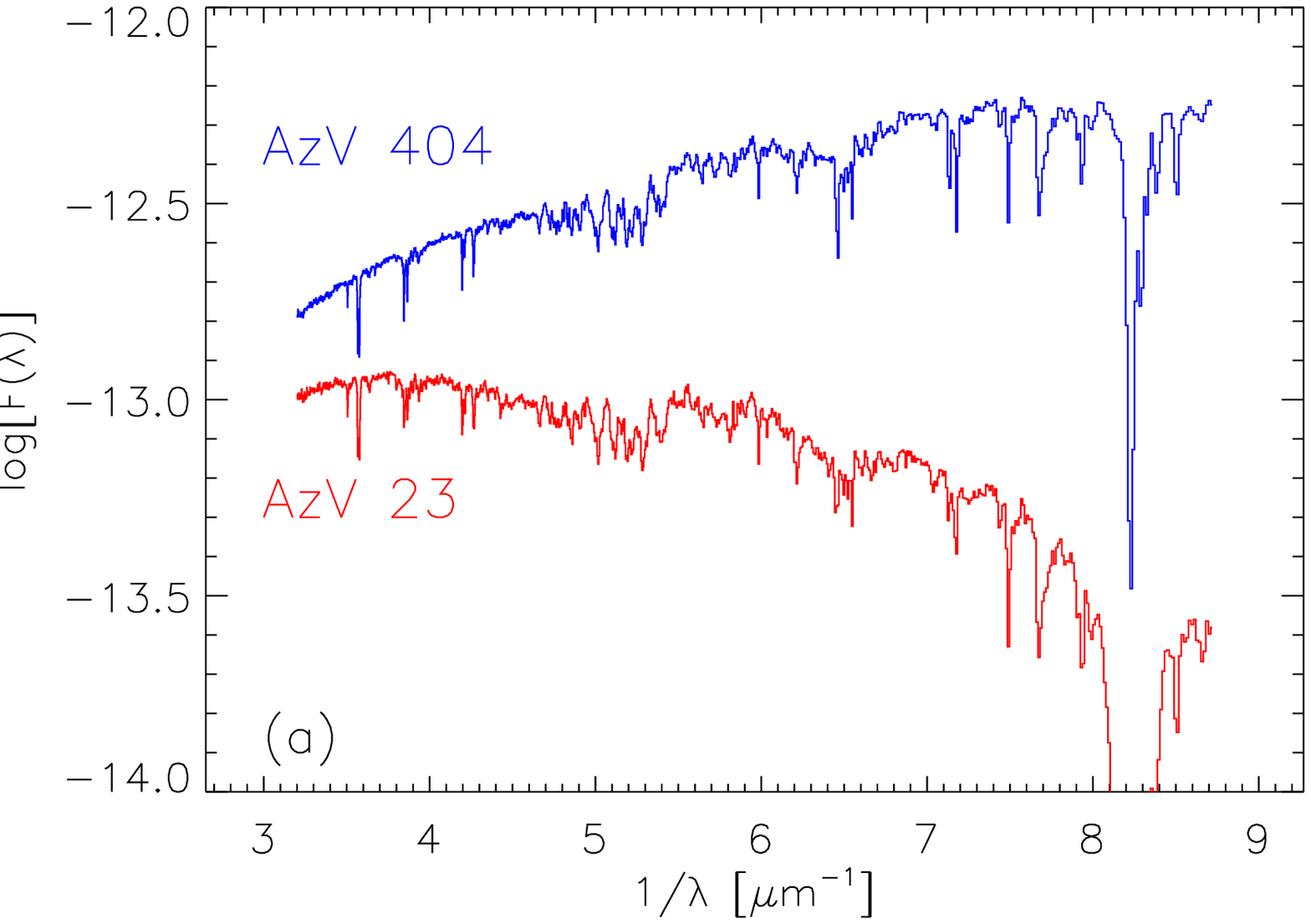}{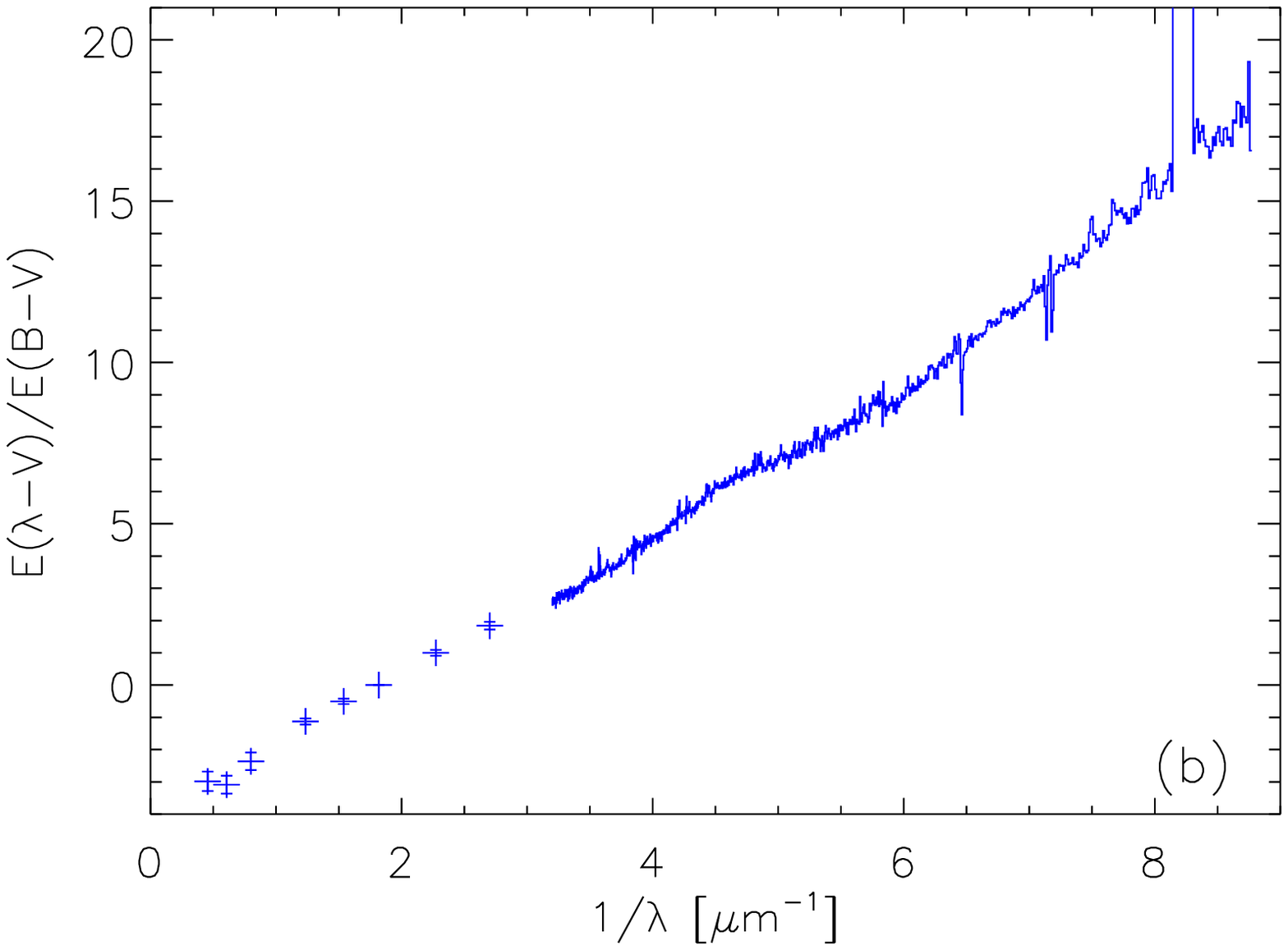}
\caption{The STIS spectra of AzV~23 and 404 are plotted in (a). The
  resulting extinction curve for AzV~23 is shown in (b).  The region
  near Ly$\alpha$ has been corrected for \ion{H}{1} absorption.
  \label{fig_stis}} 
\end{figure*}

The UV spectra for the final two stars (AzV~23 and 404) were taken
with the STIS instrument on HST as part of our GO program \#8198.  The
spectra were taken using the 52x0.5$\arcsec$ slit with the G140L and
G230L gratings.  The individual observations were coadded to produce
spectra extending from 1140 to 3140~\AA\ with a resolution of
approximately 1000.  These two spectra are presented in
Fig.~\ref{fig_stis}.  The excellent match between the reddened and
comparison spectral types can be easily seen in this figure.  In
addition, the superior nature of STIS ultraviolet spectra as compared
to IUE spectra can be seen by comparing this figure with Fig.~1 of
\citet{gor98} which displays IUE spectra for similar spectral type
stars.

\subsection{Extinction Curves \label{sec_ext_curves}}

The extinction curves for sightlines in the Magellanic Clouds were
derived using the standard pair method.  The reddened/comparison star
pairs used are listed in Table~\ref{tab_ext_data}.  We rederived the
23 extinction curves presented in \citet{gor98} and \citet{mis99} to
take into account the new calibration of IUE low-dispersion data
\citep{mas00} and the new optical and near-infrared photometry.  The
effects of the new calibration of the IUE spectra on the UV extinction
curves was small, mainly reducing the noise in the curves.  The one
new extinction curve in the SMC for the AzV~23 sightline is presented
in Fig.~\ref{fig_stis} at it's full spectral resolution.

\begin{figure*}[tbp]
\plotone{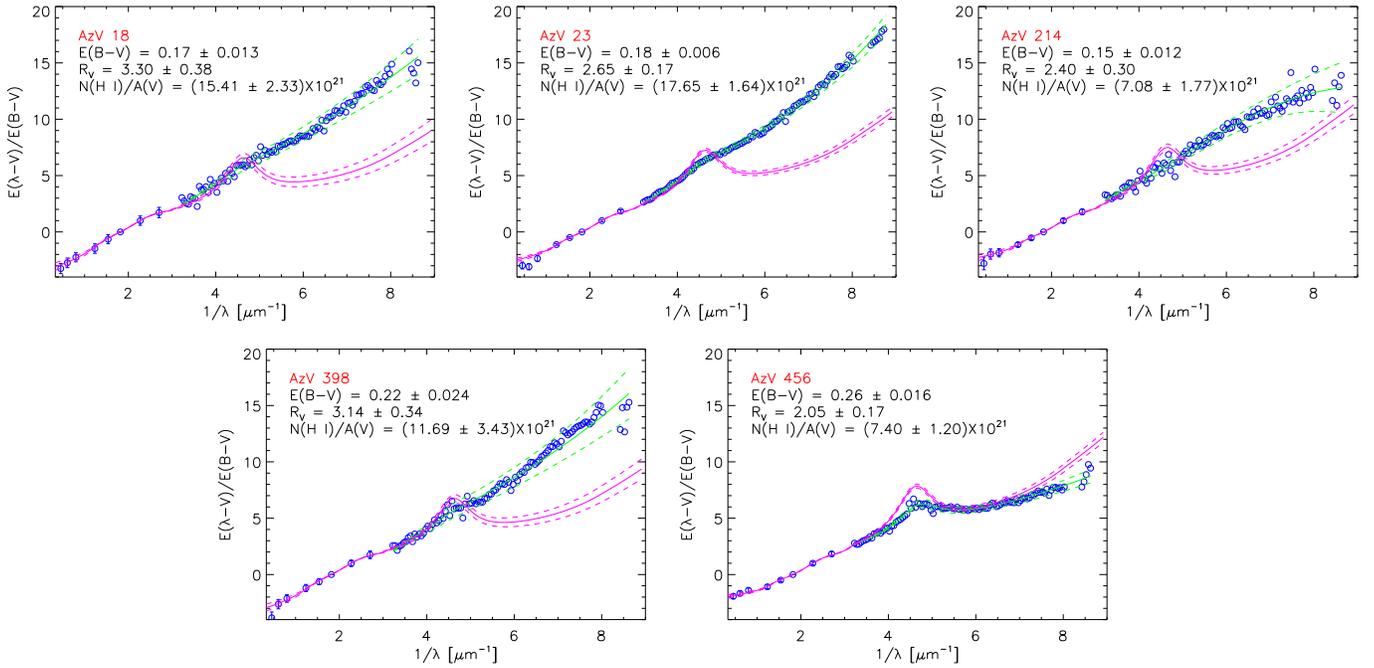}
\caption{The SMC extinction curves are plotted.  The best fit FM90
curve (solid line) and its uncertainties are plotted (dashed lines).
The CCM curve for the measured $R_V$ value (dotted line) and its
uncertainties (dot-dashed lines) are plotted.
\label{fig_smc_curves}}
\end{figure*}

\begin{figure*}[tbp]
\plotone{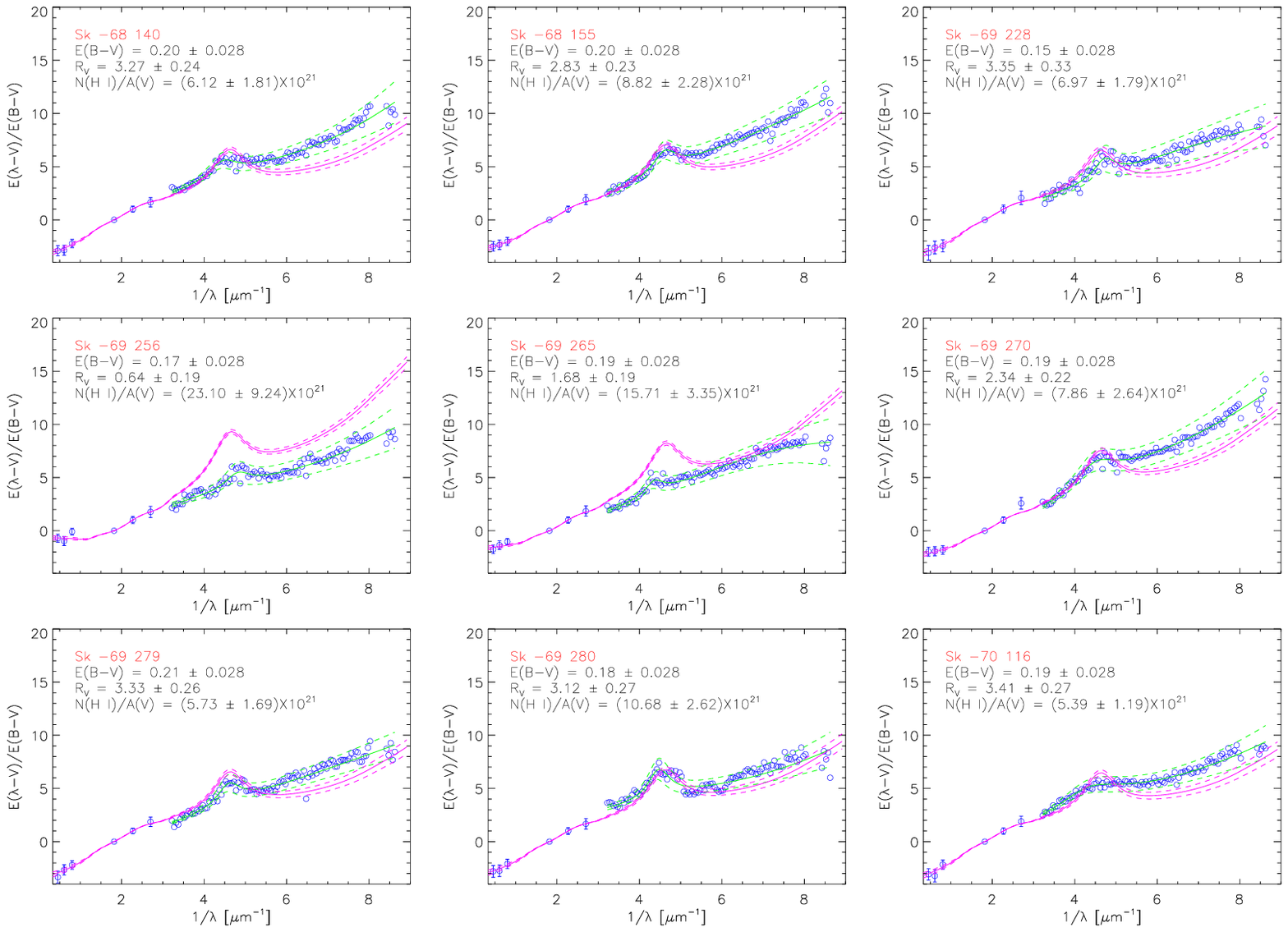}
\caption{The LMC extinction curves in the LMC-2 sample are plotted.
The best fit FM90 curve (solid line) and its uncertainties are plotted
(dashed lines).  The CCM curve for the measured $R_V$ value (dotted
line) and its uncertainties (dot-dashed lines) are plotted.
\label{fig_lmc_lmc2_curves}}
\end{figure*}

\begin{figure*}[tbp]
\plotone{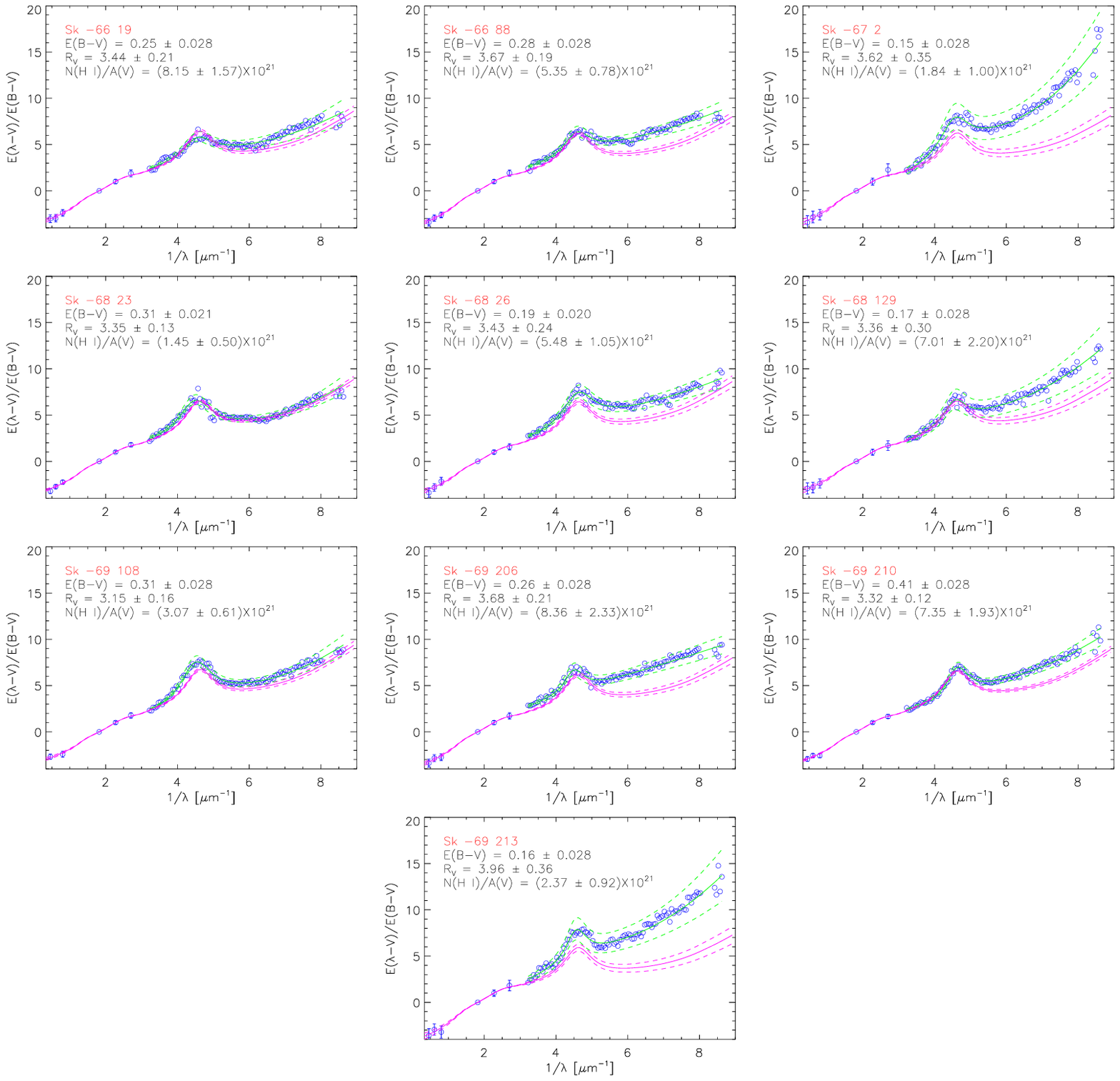}
\caption{The LMC extinction curves in the LMC-average sample are
plotted.  The best fit FM90 curve (solid line) and its uncertainties
are plotted (dashed lines).  The CCM curve for the measured $R_V$
value (dotted line) and its uncertainties (dot-dashed lines) are
plotted.
\label{fig_lmc_gen_curves}}
\end{figure*}

We calculated all 24 extinction curves and their associated
uncertainties as outlined in \citet{gor98}.  In addition, we removed
the effects of the Ly$\alpha$ \ion{H}{1} absorption using our
measurements of the \ion{H}{1} column (see \S\ref{sec_hi}).  These
extinction curves give the difference in extinction between the two
sightlines to the reddened and comparison stars.  The Milky Way
foreground component is effectively removed as long as the reddening
due to this component towards the reddened and comparison star pairs
are similar \citep{mis99}.  This results in an extinction curve that only
measures dust in the Magellanic Clouds.  All 24 extinction curves are
plotted in Figs.~\ref{fig_smc_curves}-\ref{fig_lmc_gen_curves}.

\subsection{FM Parameters}

We fit each curve with the FM parameterization of the shape of the UV
extinction curve \citep{fit90}.  The FM parameterization is
\begin{equation}
E(x - V)/E(B-V) = C_1 + C_2x + C_3D(x,\gamma,x_o) + C_4F(x)
\label{fm_eq}
\end{equation}
where $x = \lambda^{-1}$,
\begin{equation}
D(x,\gamma,x_o) = \frac{x^2}{(x^2 - x_o^2)^2 + x^2 \gamma^2},
\end{equation}
and
\begin{equation}
F(x) = 0.5392(x - 5.9)^2 + 0.05644(x - 5.9)^3
\end{equation}
for $x \geq 5.9$ and $F(x) = 0$ for $x < 5.9$.  We determined the FM
parameters for the extinction curves by numerically minimizing the
$\chi^2$ in a manner similar to that used by \citet{fit90}.  First,
$x_o$ and $\gamma$ were fixed and the values of $C_1$, $C_2$, $C_3$,
and $C_4$ were determined by minimizing the $\chi^2$.  Next, $C_1$,
$C_2$, $C_3$, $C_4$, and $\gamma$ were fixed and the value of $x_o$
was determined by minimizing the $\chi^2$.  Finally, $C_1$, $C_2$,
$C_3$, $C_4$, and $x_o$ were fixed and the value of $\gamma$ was
determined by minimizing the $\chi^2$.  These two steps were repeated
until the $\chi^2$ no longer changed by a significant amount.  This
three step method gives a smaller $\chi^2$ than doing a single
$\chi^2$ minimization while simultaneously fitting all 6 parameters.
It is worth noting that there are probably only 5 independent
parameters in the FM equation.  While all evidence points to $C_1$ and
$C_2$ being correlated in the Milky Way and the Magellanic Clouds
\citep{fit88, mis99}, we have not assumed this correlation.

For the 4 SMC Bar extinction curves, we set $x_o = 4.6$ and $\gamma =
1.0$ as the weak to nonexistent 2175~\AA\ bumps in these curves
precludes fitting all three bump parameters.  This allows for a more
realistic measurement of the bump strength or an upper limit for these
weak bump sightlines.

The uncertainties in the FM parameters for low $E(B-V)$ sightlines are
dominated by the uncertainty in $E(B-V)$ and the random uncertainties
of each wavelength point.  The uncertainties in the extinction curves
were calculated using eq.~2 of \citet{gor98}.  From this equation, it
can be seen that an uncertainty in $E(B-V)$ affects the extinction
curve in a correlated way.  Uncertainties in $E(B-V)$ do not add
random noise to each wavelength point, but shift the entire extinction
curve up and down.  The random uncertainties at each wavelength point
contribute a smaller uncertainty, except in the case of $x_o$ and
$\gamma$.  The contribution of uncertainties in $E(B-V)$
($\sigma[E(B-V)]$) to the FM
parameter uncertainties were calculated by fitting the two additional
curves for each measured extinction curve which describe the effects
of the $E(B-V)$ uncertainty.  These curves are $\left( 1 + \sigma
E(B-V)/E(B-V) \right) E(x - V)/E(B-V)$ and $\left( 1 - \sigma
E(B-V)/E(B-V) \right) E(x - V)/E(B-V)$.  The $E(B-V)$ uncertainties in
each FM parameter were then one half the difference between the
parameters for these two curves.

The FM parameter uncertainties due to the random uncertainties
($\sigma(random)$) were calculated using a Monte Carlo approach.
Determining the random component of the FM parameter uncertainties is
not usually done, but we found that the random component dominates the
$x_o$ and $\gamma$ uncertainties.  This method consisted of generating
100,000 possible FM fits similar to the best FM fit and determining
which ones fit equally well within $3\sigma$ using the F-test where
the $\sigma$ is only that due to random flux uncertainties.  The
resulting $1\sigma(random)$ uncertainties were determined by dividing
by three.  The reported FM uncertainties (see
Table~\ref{tab_fm_param}) were determined by summing in quadrature the
$E(B-V)$ associated ($\sigma[E(B-V)]$) and random ($\sigma(random)$)
uncertainties.  The $E(B-V)$ uncertainties dominate the $C_1$, $C_2$,
$C_3$, and $C_4$ uncertainties, while the random uncertainties
dominate the $x_o$ and $\gamma$ uncertainties.

\subsection{$R_V$ values}

The value of $R_V = A_V/E(B-V)$ for each sightline was determined
using a $\chi^2$ minimization method.  This method relies on the
invariance of the RIJHK portion of the extinction curve \citep{rie85,
mar90}.  The $R_V$ value for each extinction curve is the value which
minimizes the $\chi^2$ between the measured RIJHK extinction and the
\citet{rie85} curve.  The equation giving $\partial \chi^2 / \partial
R_V = 0$ was solved and the analytic solution for $R_V$ was found.
From this equation, the uncertainty for $R_V$ was derived.  See the
appendix for details of this derivation.  The $R_V$ values and
uncertainties are tabulated in Table~\ref{tab_ext_data}.  Our method
of using the RIJHK extinction curve itself to determine $R_V$ values
is different from what is usually done.  Usually the observed VRIJHK
or just VK photometry of the reddened star is compared to assumed
intrinsic colors \citep{mor82, car89, gor98, mis99}.  In the
Magellanic Clouds, the reddened stars are usually first corrected for
the average Milky Way foreground reddening \citep{mor82}.  Instead, we
have used the measured VRIJKH photometry for the matched comparison
star and avoided having to assume intrinsic colors of the reddened
star.  This naturally removes the foreground Milky Way reddening and
ensures that the measured $R_V$ value corresponds to the same dust
column as the extinction curve.  The CCM relationship for these $R_V$
values for each measured extinction curve is given in
Figs.~\ref{fig_smc_curves}-\ref{fig_lmc_gen_curves}.  The effects of
the uncertainty in $R_V$ on the CCM relationship is shown in these
plots using dashed curves.

\subsection{\ion{H}{1} columns \label{sec_hi}}

\begin{figure*}[tbp]
\epsscale{2.1} 
\plottwo{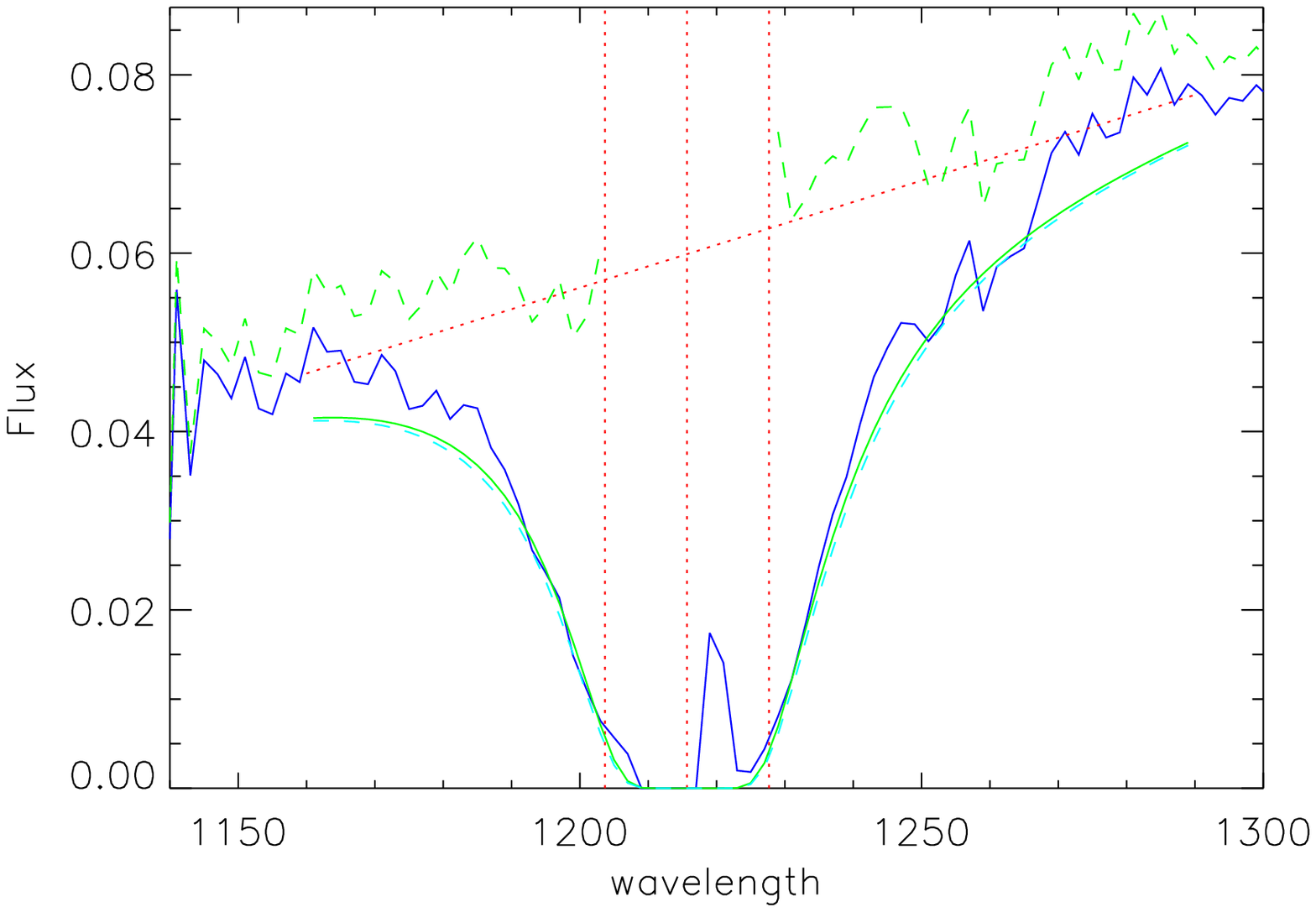}{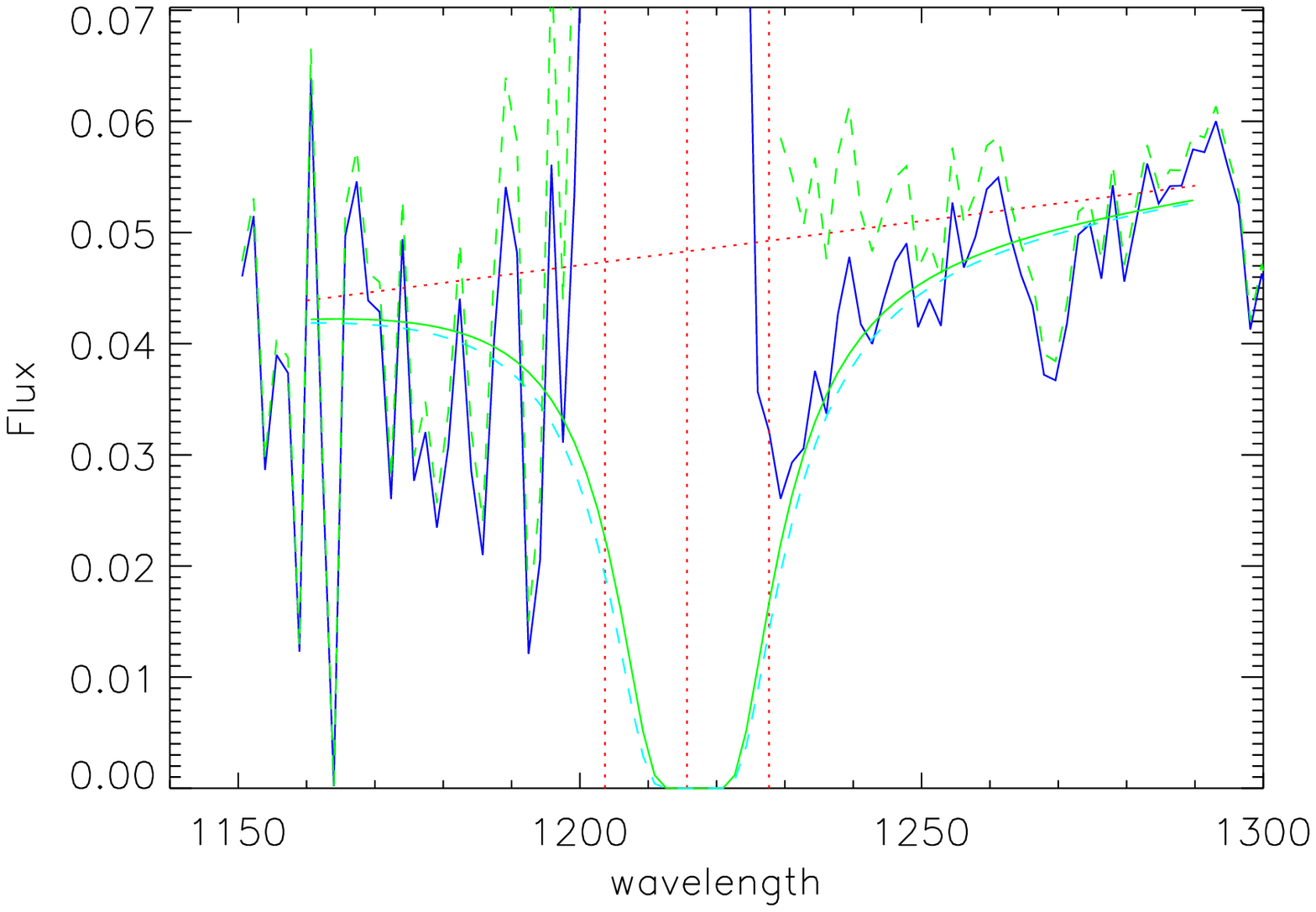}
\caption{The ratio spectra for the AzV 23/404 pair (a) and
Sk~-69~108/Sk~-67~78 pair (b) are shown (solid line) with the best fit
\ion{H}{1} profile (dot-dashed line).  The ratio spectrum divided by
the best fit model Ly$\alpha$ profile is shown as a dashed line.  The
vertical dotted lines give the center of the Ly$\alpha$ line for the
SMC (a) and LMC (b) velocities and $\pm 12$\AA\ region which is
excluded from the fit.  The dotted line gives a nominal continuum.
\label{fig_hi_measure}}
\end{figure*}

We determined the \ion{H}{1} column for each extinction curve using a
variant of the method outlined in \citet{boh75}.  This method relies
on fitting the wings of the Ly$\alpha$ absorption profile.  As we are
only interested in a measurement of the difference in the \ion{H}{1}
column between the reddened and comparison stars, we made this
measurement in the ratio spectrum for each reddened and comparison
star pair.  The uncertainty in this measurement was estimated by eye,
varying the \ion{H}{1} value until it was noticeable incorrect.
Two examples of this are shown in Fig.~\ref{fig_hi_measure}.
Basically, the strength of a model Ly$\alpha$ profile is adjusted
until the division of the ratio spectrum by the model spectrum yields
a straight line.  The central wavelength of the Ly$\alpha$ line for
the LMC and SMC is corrected for the heliocentric velocity of the LMC
and SMC (280 and 130 km/s, respectively).  The central region of the
Ly$\alpha$ line is ignored as it is contaminated by geocoronal
emission.  This measurement is straightforward for the STIS data.  For
IUE data, this measurement is much more uncertain as the contamination
by geocoronal emission is much larger and the spectra blueward of
Ly$\alpha$ are very noisy.  The red wing of Ly$\alpha$ was mainly used
in the measurement of \ion{H}{1} columns for the IUE data.  The
measured \ion{H}{1} values and uncertainties are tabulated in
Table~\ref{tab_ext_data}.

\section{Discussion \label{sec_discuss}}

\subsection{Comparison of Magellanic Cloud and Milky Way Extinction
Curves}

Now that we have produced full ultraviolet to near-infrared extinction
curves (in units of $A(\lambda)/A(V)$) for all the known Magellanic
Cloud reddened/comparison star pairs, we can quantitatively compare
them to Milky Way extinction curves.  The $R_V$ dependent
CCM relationship and the CCM sample extinction curves provides a nice,
compact form for representing the properties of Milky Way dust found
in the local interstellar medium.  Thus, we can compare Milky Way and
Magellanic Clouds dust by seeing if the CCM relationship is
applicable to any of the individual extinction curves or describes
aspects of the sample behavior of these curves.

To test if the CCM relationship accurately describes any of the
individual extinction curves, we compare the full measured UV to NIR
extinction curves to the appropriate CCM curve for the measured $R_V$.
This is done in Figs.~\ref{fig_smc_curves}-\ref{fig_lmc_gen_curves}
and includes curves showing the uncertainty in the measured curve as
well as the uncertainty in the CCM curve due to uncertainty in the
measured $R_V$ value.  It should be remembered that the CCM
relationship gives the average behavior at a particular $R_V$ and that
individual curves can have small deviations from this average behavior
and still follow the CCM relationship.

It is clear from Fig.~\ref{fig_smc_curves} that the four curves
located in the star-forming bar of the SMC (AzV 18, 23, 214, \& 398)
do not follow the CCM relationship.  For the one curve located outside
of the SMC bar (AzV 456), the CCM curve does a better job following
the measured curve, but there still are significant deviations at the
2175~\AA\ bump and in the far-UV ($> 7 \micron^{-1}$).  Thus, there
are no measured extinction curves in the SMC which follow the CCM
relationship.

In the LMC, there is one curve in the LMC2 sample (Sk~-69~280) and
three curves in the LMC-average sample (Sk~-66~19, Sk~-68~23, \&
Sk~-69~108) which follow the CCM relationship within their
uncertainties
(Figs.~\ref{fig_lmc_lmc2_curves}-\ref{fig_lmc_gen_curves}).  Thus, 
there is evidence that the CCM relationship is at work in the LMC, but
only in a limited sense as it describes only 4 out of the 19 measured
extinction curves.

\begin{figure*}[tbp]
\plotone{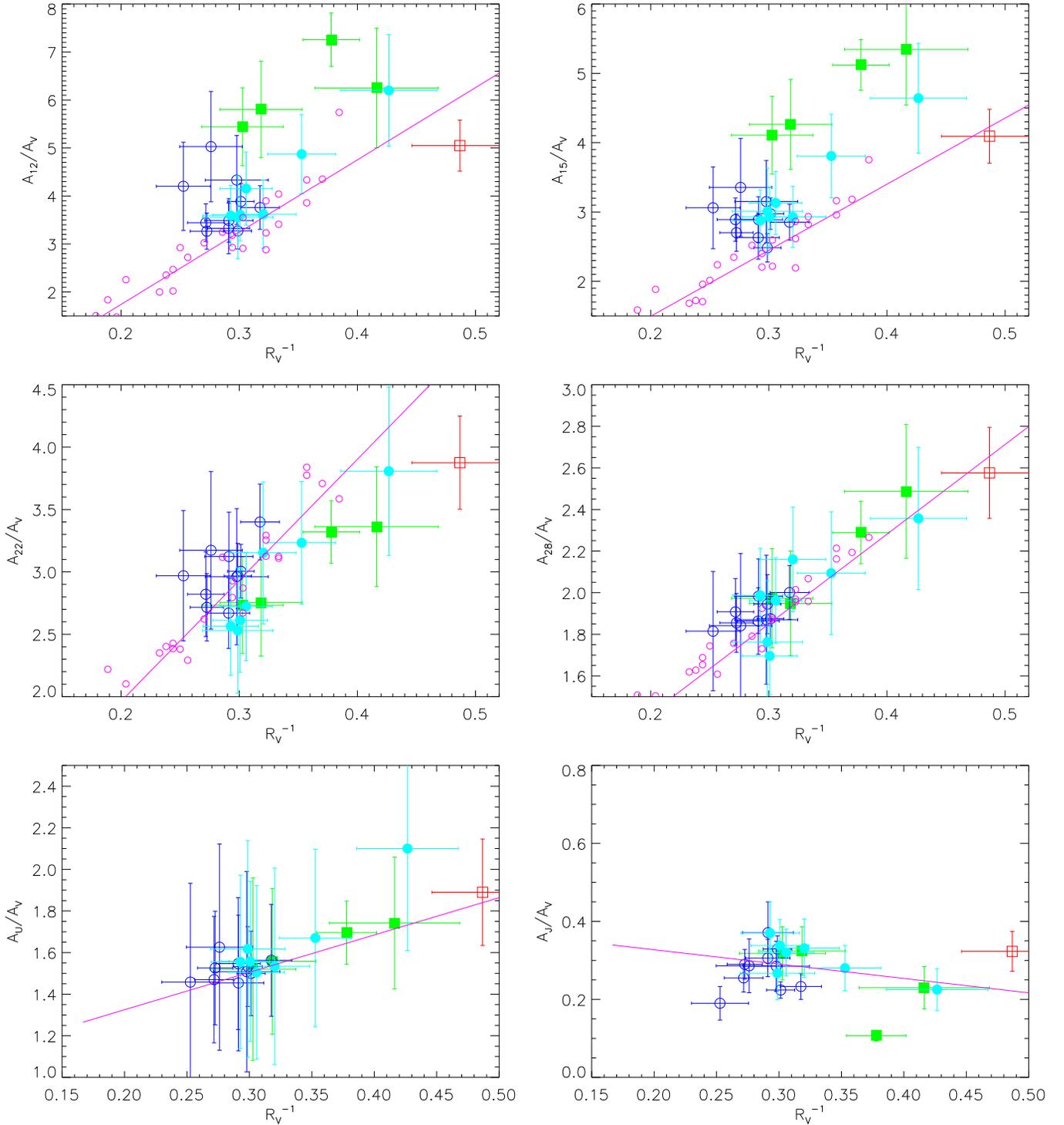}
\caption{Plots of $A_{\lambda}/A_V$ versus $R_V^{-1}$ are shown where
$A_{12} = A(1200$~\AA), $A_{15} = A(1500$~\AA), $A_{22} =
A(2200$~\AA), $A_{28} = A(2800$~\AA), $A_U = A(3500$~\AA, U band),
$A_J = A(1.25$~\micron, J band).  The LMC and SMC measurements are
given by the circles (solid = Average and open = LMC2) and squares
(solid = Bar and open = Wing) with error bars, respectively.  The line
gives the CCM relationship and the open circles without error bars the
original \citet{fit90} data used by CCM ($\lambda < 3000$~\AA\ plots
only).  \label{fig_al_rv}}
\end{figure*}

The most direct way to quantify how applicable the CCM relationship is
in the Magellanic Clouds is to plot $R_V^{-1}$ versus
$A_{\lambda}/A_V$.  This was how the CCM relationship was originally
presented by \citet{car89}.  Fig.~\ref{fig_al_rv} shows such plots for
$\lambda =$ 1200, 1500, 2200, \& 2800~\AA\ and U and J bands.  Since
the CCM relationship represents the average extinction behavior as a
function of $R_V$, saying extinction measurements do not follow CCM is
to say that they are beyond the scatter of the extinction curves which
were used to derive the CCM relationship.  In Fig.~\ref{fig_al_rv},
the small open circles without error bars give data for the extinction
curves used to derive CCM and can be used to determine the scatter
which is consistent with the CCM relationship.  These plots show that
at far-UV wavelengths ($\lambda =$ 1200 \& 1500~\AA) the CCM
relationship forms a lower bound to the values of $A(\lambda)/A(V)$ at
a particular value of $R_V$.  At longer wavelengths ($\lambda \ge
2200$~\AA), the Magellanic Clouds measurements are indistinguishable
from the CCM relationship within their uncertainties.

\begin{figure*}[tbp]
\plotone{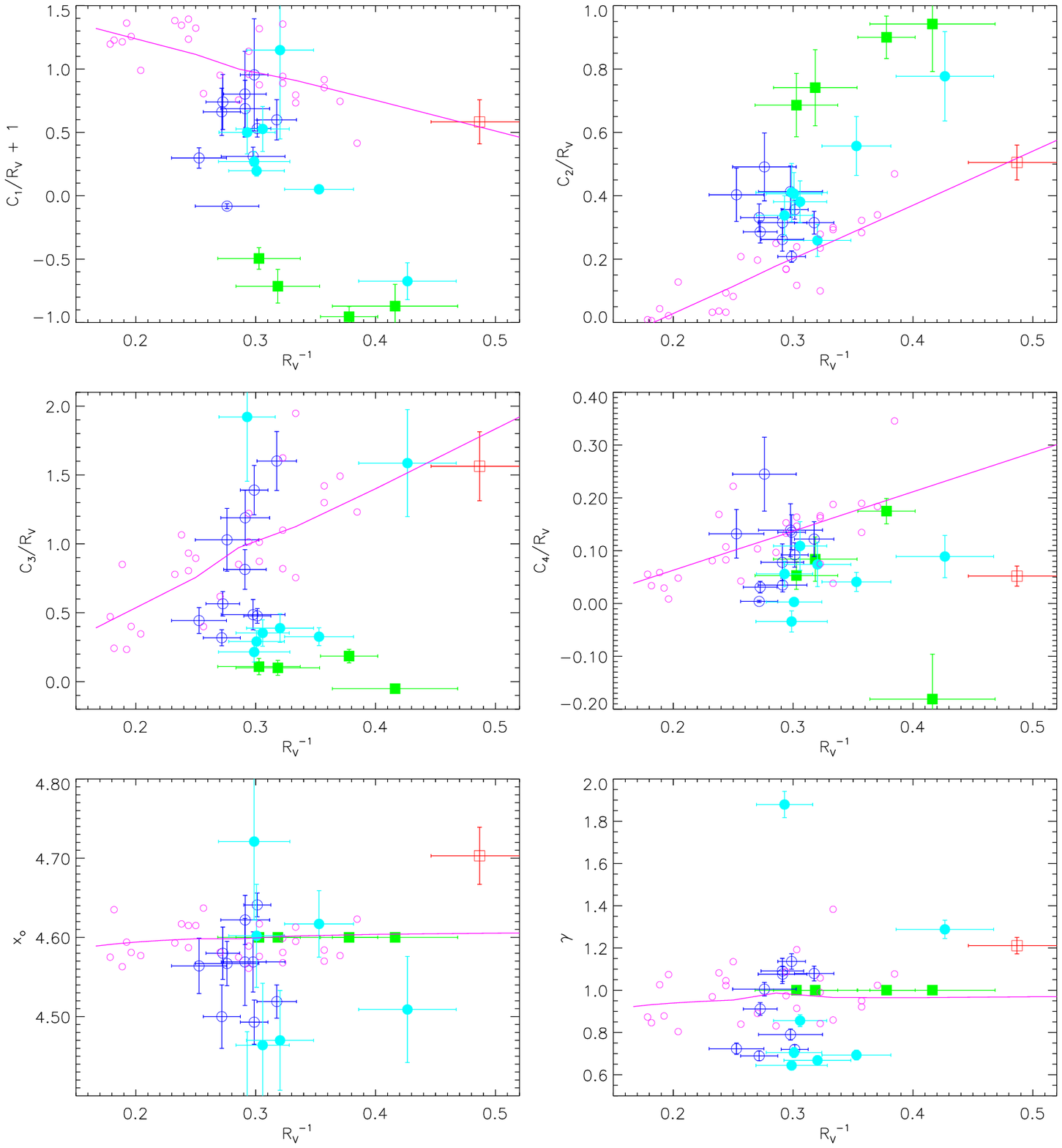}
\caption{The FM fit parameters are plotted versus $R_V^{-1}$.  The LMC
and SMC measurements are given by the circles (solid = Average and
open = LMC2) and squares (solid = Bar and open = Wing) with error
bars, respectively.  The line gives the CCM relationship and the open
circles without error bars the original \citet{fit90} data used by
CCM. \label{fig_fm_rv}}
\end{figure*}

A final way to probe how well the CCM relationship works in the
Magellanic Clouds is to examine behavior of the FM parameters as a
function of $R_V$.  The FM parameters describe the UV extinction curve
with only 6 parameters allowing for more sensitive tests to be
performed.  The FM parameters $C_1$, $C_2$, $C_3$, and $C_4$ have a
factor of $R_V$ embedded in them, we have plotted the equivalent
$R_V$-independent coefficients versus $R_V^{-1}$ in
Fig.~\ref{fig_fm_rv}.  These $R_V$-independent coefficients can be
derived by examining eq.~\ref{fm_eq} converted from $E(x-V)/E(B-V)$ to
$A(x)/A(V)$ units.  The equation expressed this way is
\begin{eqnarray}
A(x)/A(V) & = & \frac{E(x - V)}{E(B-V)}\frac{1}{R_V} + 1 \\
          & = & (C_1/R_V + 1) + (C_2/R_V)x + (C_3/R_V)D(x,\gamma,x_o) 
  + (C_4/R_V)F(x).
\end{eqnarray}
The behavior of the $(C_1/R_V + 1)$, $C_2/R_V$, $C_3/R_V$, and
$C_4/R_V$ coefficients do not follow the CCM relationship directly,
but the CCM relationship does form a bound on the values of these
coefficients as a function of $R_V$.  For example, all the values of
$C_2/R_V$ lie on or above the line defining the CCM relationship.  In
the plots of the other three coefficients, the CCM relationship forms
an upper bound on their behaviors.  In the case of $x_o$ and $\gamma$,
CCM does not predict much of a dependence on $R_V$ and we do not see
one for the Magellanic Clouds either.  The values of $x_o$ and
$\gamma$ for the Magellanic Clouds have a much larger scatter than
seen in the extinction curves defining CCM.  For $x_o$ this scatter is
consistent with the uncertainties in $x_o$.  For $\gamma$, the scatter
is larger than can be accounted for by measurement uncertainties.
Thus, it is either real or the result of measurement uncertainties we
have not accounted for in our error analysis.

By examining the behavior of the Magellanic Cloud extinction curves
with $R_V$ three different ways and comparing that behavior to that
predicted by the $R_V$ dependent CCM relationship, we find evidence
that the general behavior of Milky Way extinction curves is seen in
the Magellanic Clouds.  Not only are 4 LMC extinction curves
indistinguishable from Milky Way extinction curves, but the general
behavior of Milky Way extinction curves forms a bound on the general
behavior of Magellanic Cloud extinction curves.

\subsection{Super-CCM relationship? \label{sec_sccm}}

While a small number of Magellanic Cloud extinction curves do seem to
be well described by the CCM relationship, the majority do not.  Yet
there is strong evidence that the CCM relationship serves as a bound
on the behavior of all the Magellanic Cloud extinction curves.  This
fact raises the question: Is there some more general relationship
dependent on $R_V$ and at least one other parameter which describes
the average behavior of the Milky Way and Magellanic Cloud extinction
curves?

\begin{figure*}[tbp]
\plotone{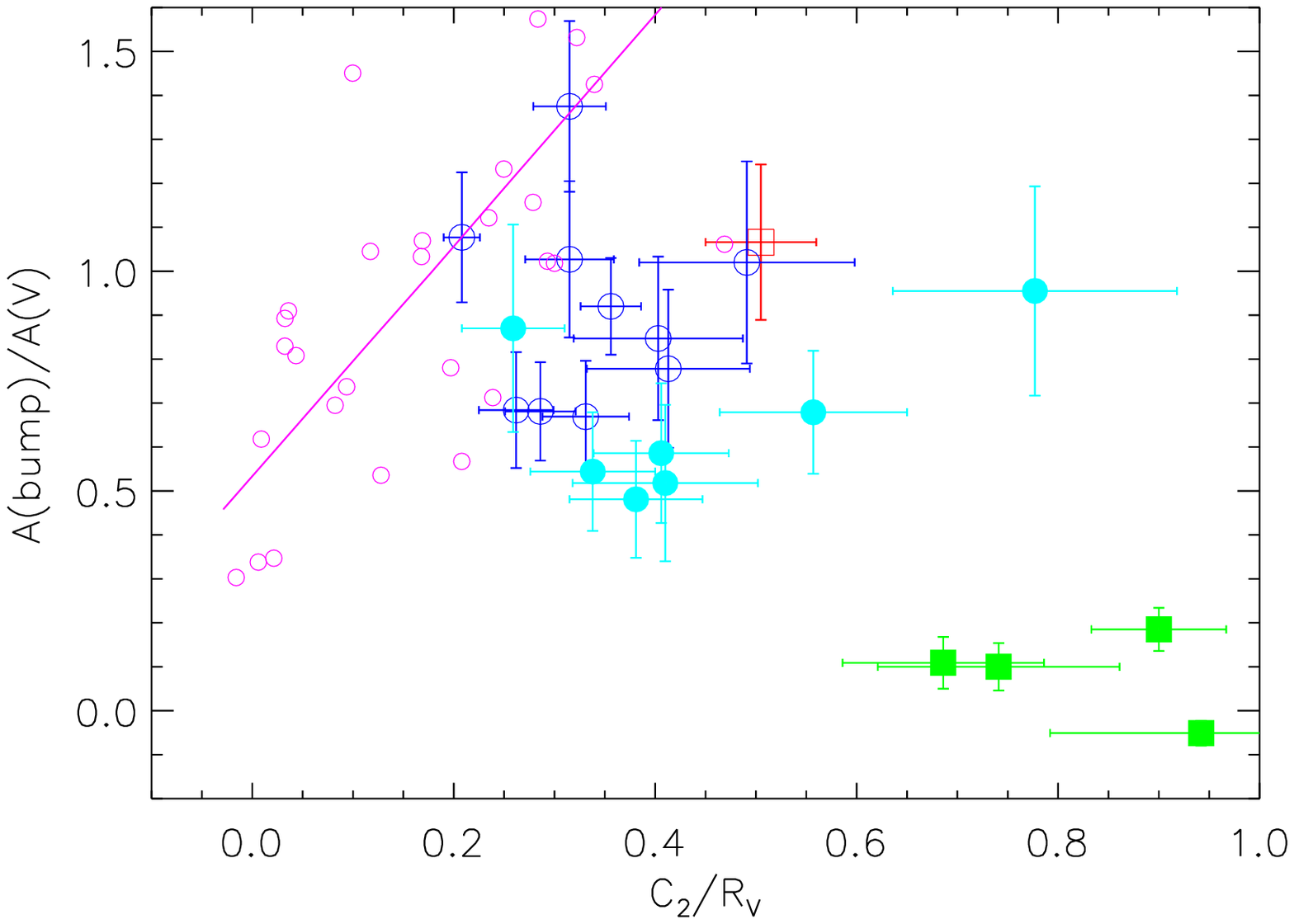}
\caption{The strength of the 2175~\AA\ bump ($A(bump)/A(V) =
C_3/(\gamma^2 R_V)$) is plotted versus the steepness of the
ultraviolet extinction ($C_2/R_V$).  The LMC and SMC measurements are
given by the circles (solid = Average and open = LMC2) and squares
(solid = Bar and open = Wing) with error bars, respectively.  The line
gives the CCM relationship and the open circles without error bars the
original \citet{fit90} data used by CCM.  \label{fig_bump_c2}}
\end{figure*}

An indication that such a relationship might exist was presented by
\citet{cla00} for low density sightlines in the Milky Way which
display some of the same deviations from CCM seen in the Magellanic
Clouds.  They found that the strength of the 2175~\AA\ bump ($(\pi
C_3)/(2\gamma)$) and the steepness of the ultraviolet extinction
($C_2$) were anti-correlated along the low density sightlines as well
as for the average curves in the Magellanic Clouds \citep{gor98,
mis99}.  We present a similar plot in Fig.~\ref{fig_bump_c2} for the
Magellanic Cloud extinction curves with the equivalent $R_V$
independent measures of the 2175~\AA\ bump strength and ultraviolet
extinction steepness.  This plot gives evidence for an
anti-correlation between the strength of the 2175~\AA\ bump and the
steepness of the ultraviolet extinction.

\begin{figure*}[tbp]
\epsscale{2.1} 
\plottwo{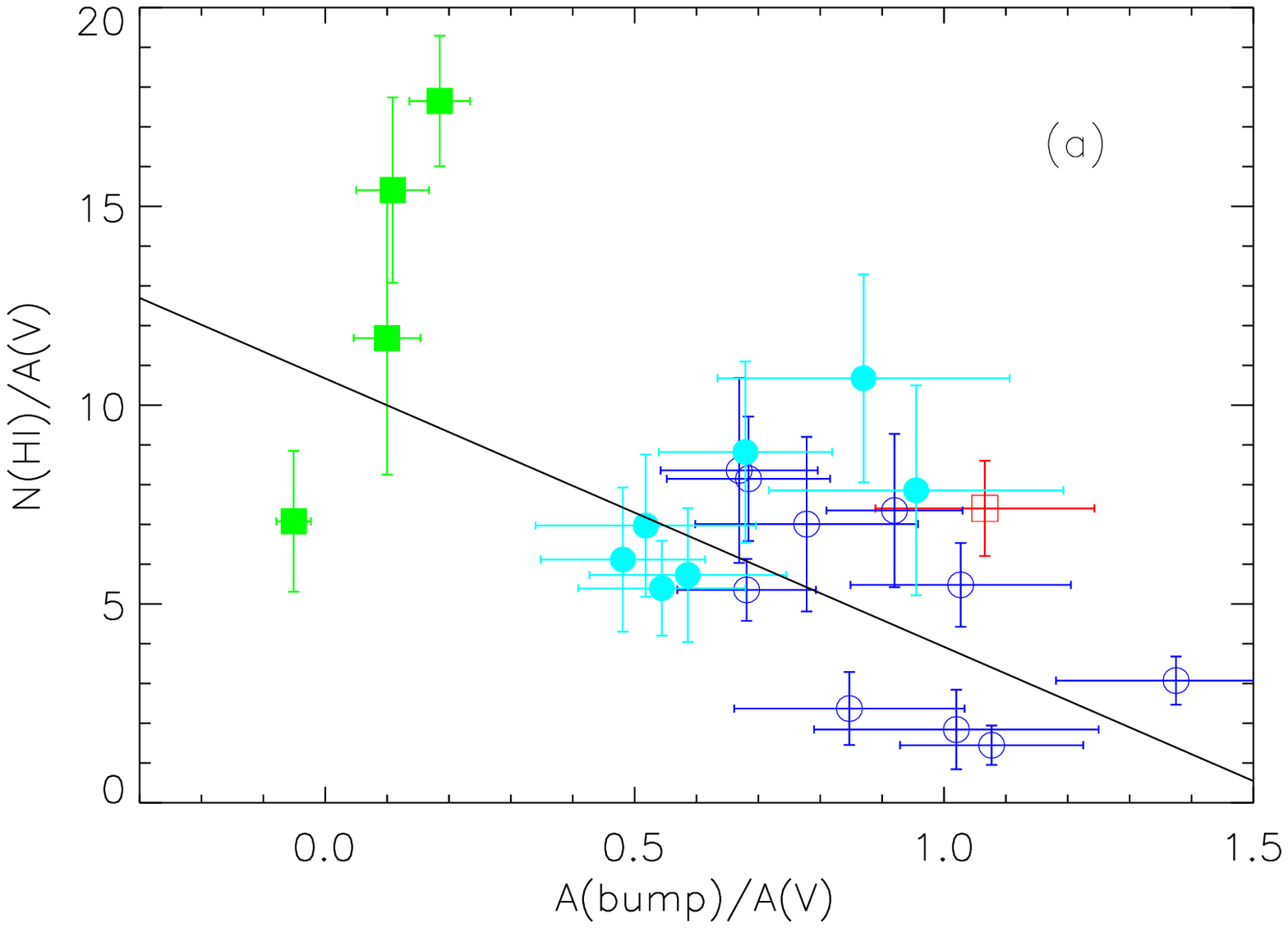}{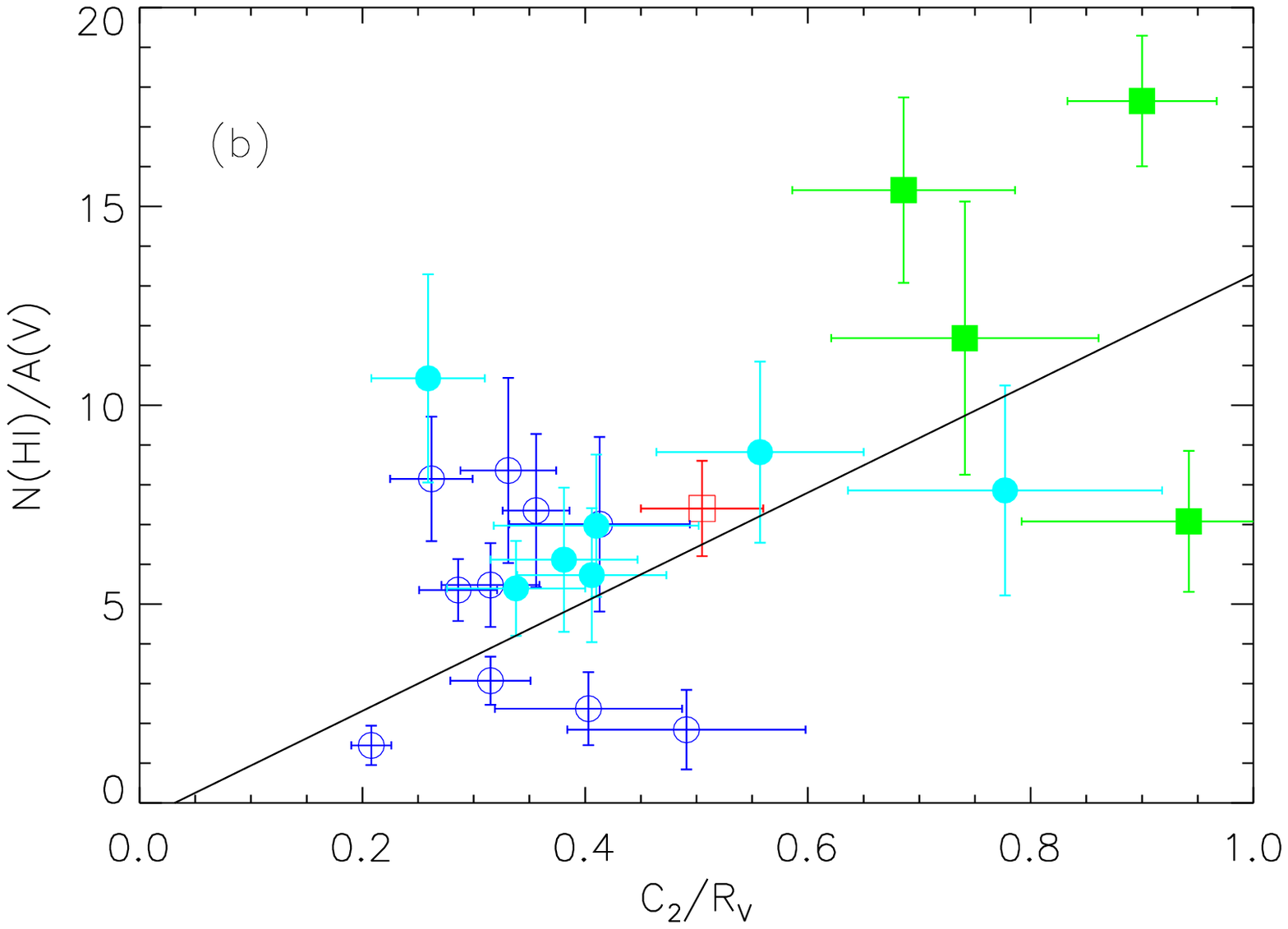}
\caption{The behavior of the gas-to-dust ratios ($N(HI)/A(V)$) versus
(a) 2175~\AA\ bump strength ($A(bump)/A(V)$) and (b) ultraviolet
steepness ($C_2/R_V$) is shown.  The LMC and SMC measurements are
given by the circles (solid = Average and open = LMC2) and squares
(solid = Bar and open = Wing) with error bars, respectively.  The line
gives a simple linear fit to the data. \label{fig_hiav}}
\end{figure*}

One possible second parameter could be the measured gas-to-dust ratio
($N(HI)/A(V)$).  To test this we have plotted in Fig.~\ref{fig_hiav}
the gas-to-dust ratios versus the 2175~\AA\ bump strength and
ultraviolet steepness values for the Magellanic Cloud extinction
curves.  The scatter in these two plots is quite large as are the
uncertainties on the individual points.  There might be real
correlations in both plots, but higher quality data are needed.  There
would be reason to expect a correlation between the gas-to-dust ratio
and the behavior of extinction curves.  The gas-to-dust ratio is known
to correlate with metallicity on a galaxy wide basis \citep{iss90}.
On a local scale, it could be a measure of the formation and
destruction history of dust grains.  For example, the dust
self-shielding will decrease with increasing gas-to-dust ratio making
it easier for dust grains to be destroyed by the ambient radiation
field.  If the dust grains responsible for the 2175~\AA\ bump are
easier to destroy than those responsible for the underlying
ultraviolet extinction, a behavior like that seen in
Fig.~\ref{fig_hiav} would be expected.

\subsection{SMC Bar extinction curves and the 2175~\AA\ bump}

The extremely weak or absent 2175~\AA\ bump in the four SMC Bar
extinction curves makes these curves unique.  In all other measured
extinction curves, the 2175~\AA\ bump is quite prominent.  The obvious
question is whether the weak 2175~\AA\ bump in the SMC Bar is due to
Milky Way contamination or is intrinsic to the SMC.  While only one of
the four curves has a bump which is detected at greater than
$3\sigma$, the other three are all $2\sigma$ detections.  The
construction of the curves using SMC comparison stars should remove
all the Milky Way foreground, but small differences between the
reddened and comparison stars foreground could result in a weak
2175~\AA\ bump.  A foreground contamination like this would result in
a positive bump as often as a negative bump.  Interestingly, one out
of the four SMC Bump curves (AzV~214) has a $2\sigma$ detection of a
negative bump and a negative far-UV curvature ($C_4$).  The foreground
contamination (in percent of total $E(B-V)$ with $1\sigma$
uncertainties) needed to produce spurious bump detections at the
observed levels would have to be ($25 \pm 14$)\%, ($39 \pm 10$)\%,
($10 \pm 5$)\%, and ($23 \pm 13$)\% for AzV~18, 23, 214, and 398,
respectively.  These foreground contaminations are not unlikely given
the low $E(B-V)$ values for these sightlines, especially given the
large uncertainties.  While it is difficult to definitively decide,
the current evidence suggests that the carriers of the 2175~\AA\ bump
could be completely absent from the dust in the SMC Bar sightlines.

\subsection{Sample Average Curves}

Since the Magellanic Cloud extinction curves display significantly
different shapes than that seen in the Milky Way, average extinction
curves of the LMC-average, LMC2-supershell, and SMC-Bar samples are of
interest.  These average curves capture the large scale variations in
dust properties at higher signal-to-noise than the individual curves.

\begin{figure*}[tbp]
\plotone{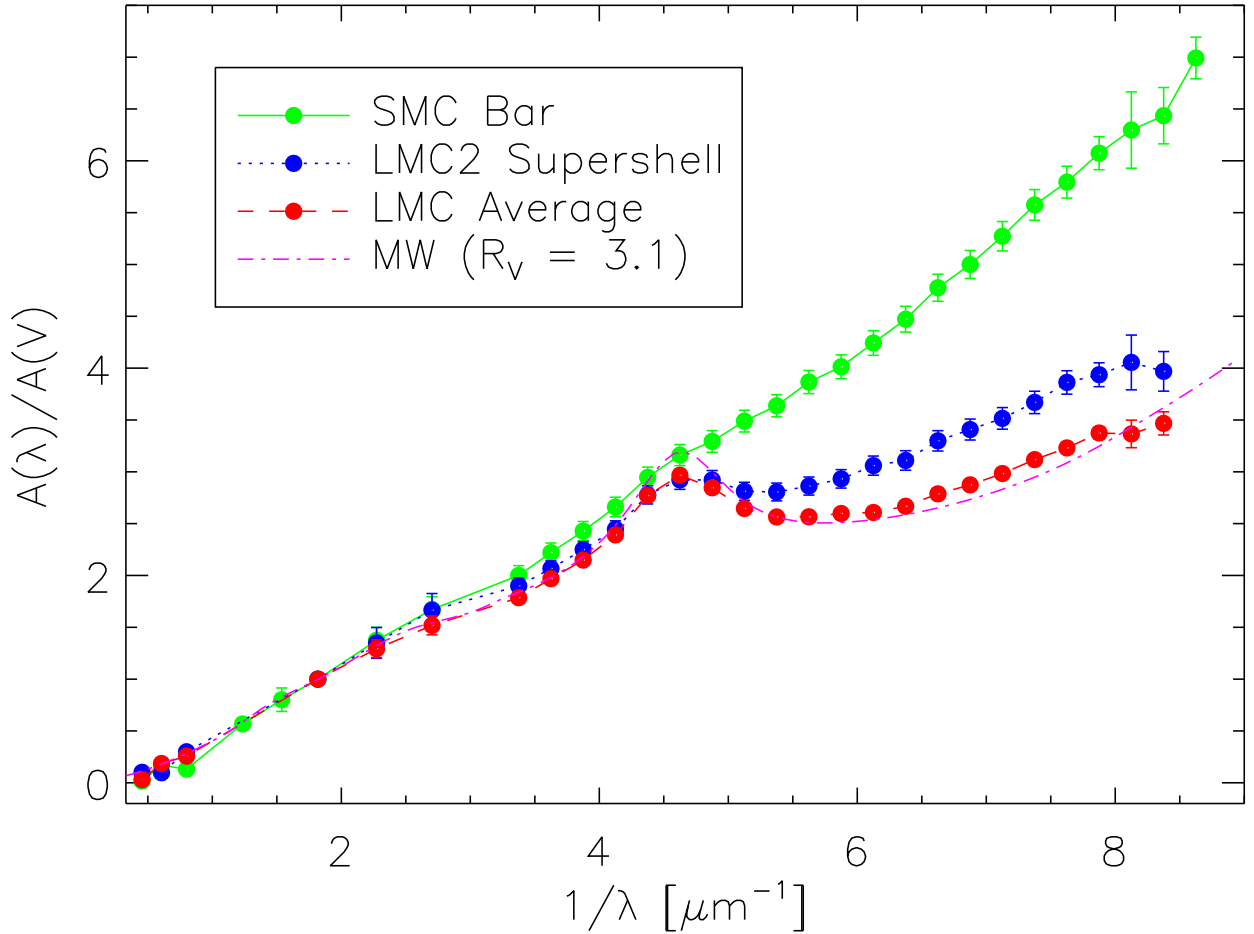}
\caption{The sample average extinction curves are plotted along with
  the ``average'' Milky Way curve (CCM with $R_V = 3.1$).
  \label{fig_ave_curves}} 
\end{figure*}

While past work has produced such average curves \citep{gor98, mis99},
we can create more accurate average curves as the result of our
determinations of $R_V$ values.  We have averaged the individual
curves in each sample in $A(\lambda)/A(V)$ units instead of
$E(\lambda-V)/E(B-V)$ units as was done in previous work.
$E(\lambda-V)/E(B-V)$ units are a relative measure of dust properties
whereas $A(\lambda)/A(V)$ units are an absolute measure of the dust
properties and, as a result, more accurately represent the effects of
dust.  For the LMC2 supershell sample, we have not used the results
for Sk~-69~256 in producing the LMC2 supershell average.  It's
extremely low $R_V$ value of $0.64$ is likely the result of poor JHK
photometry, contamination by nearby hot dust, or contamination by a
red companion.

The sample average curves are given in Figure~\ref{fig_ave_curves} and
tabulated in Table~\ref{tab_ave_curves}.  The averages were calculated
weighting by the uncertainties at each wavelength for each individual
curve and in 0.25~$\micron^{-1}$ wide bins.  The average values of
$R_V$ and N(\ion{H}{1})/A(V) were also calculated weighting by the
individual uncertainties and are tabulated in
Table~\ref{tab_ext_data}.  The average curves were fit with the FM
parameterization and the best fit 6 FM parameters are given in
Table~\ref{tab_fm_param}.  The FM fits were done for wavelengths $<
8.4$~$\micron$ as this is the bluest wavelength in common between the
IUE and STIS data.  This fitting limit only affected the SMC Bar
average.

For the SMC Bar, we find that $R_V = 2.74 \pm 0.13$ and
N(\ion{H}{1})/A(V)~$= 13.18 \pm 1.02$ which is consistent with the
results of \citet{bou85} and \citet{gor98}.  However, we do not find
that the gas-to-dust ratio for AzV~456 (Sk~143) is similar to that of
the Milky Way.  For the Milky Way, N(\ion{H}{1})/A(V)~$\sim 1.55$
\citep{boh78, dip94} and we find N(\ion{H}{1})/A(V)~$= 7.40 \pm 1.20$
for AzV~456 whereas \citet{bou85} found N(\ion{H}{1})/A(V)~$\sim 2.6$
assuming $R_V = 2.72$ (the average $R_V$ for the entire SMC).  The
differences between our measurement and \citet{bou85} can be traced to
our lower $E(B-V)$ and a lower $R_V$ value.  These differences
illustrate the difficulties of measuring such quantities for low
reddening sightlines.

For the LMC2 supershell sample, we find that $R_V = 2.76 \pm 0.09$ and
N(\ion{H}{1})/A(V)~$= 6.97 \pm 0.67$.  Our gas-to-dust ratio is
roughly consistent with results of \citet{koo82} and \citet{fit85} who
found N(\ion{H}{1})/A(V)~$\sim 6.3$ and $8.7$, respectively.  Our
value of $R_V$ is lower than that of \citet{cla85} who found $R_V \sim
3.5$ but consistent with \citet{mis99}.  For the LMC average sample,
we find that $R_V = 3.41 \pm 0.06$ and N(\ion{H}{1})/A(V)~$= 3.25 \pm
0.28$.  These values are significantly different from those for the
LMC2 supershell sample.  The different gas-to-dust ratios is in
conflict with the conclusions of \citet{fit86} who state that there are
no gas-to-dust measurable differences between the non-30~Dor and
30~Dor samples.  But, examination of their Fig.~6 shows that the
non-30~Dor points fall consistently below 30~Dor gas-to-dust ratio
implying that we are not in conflict with earlier work.  The lower
gas-to-dust ratio is especially interesting as it supports a
systematic trend of more extreme extinction curves (smaller 2175~\AA\
bump and steeper far-UV rise) with rising gas-to-dust ratio (see
\S\ref{sec_sccm}).

\section{Conclusions}

We have presented a quantitative, exhaustive comparison of all the
known Magellanic Cloud extinction curves with a representative sample
of Milky Way extinction curves.  Like previous studies
\citep{nan78, leq82, pre84, cla85, fit86, gor98, mis99} we find that
both the LMC and SMC have examples of extinction curves qualitatively
similar to those found in the Milky Way.  Unlike previous studies, we
are able to take this comparison one step further and make
quantitative comparisons as we determined $R_V$ values for all
Magellanic Cloud extinction curves.  This allows for the comparison of
Magellanic Cloud and Milky Way extinction curves to be done using
measurements based on absolute dust properties ($A(\lambda)/A(V)$)
instead of relative dust properties ($E(\lambda-V)/E(B-V)$).  The
importance of this difference is well illustrated by the work of
\citet{car89} where the comparison of Milky Way curves in
$A(\lambda)/A(V)$ units allowed for the derivation of the $R_V$
dependent CCM relationship.  We conclude that 4 extinction curves in the
LMC are {\em indistinguishable} from Milky Way extinction curves and
Milky Way extinction curves form a bound on the behavior of Magellanic
Cloud extinction curves.

The majority of the Magellanic Cloud extinction curves are
significantly different than Milky Way extinction curves and this is
likely a result of Magellanic Cloud extinction curves probing quite
different environments than the CCM sample of Milky Way extinction
curves.  The CCM sample is based on fairly quiescent sightlines and
most of our extinction curve measurements in the Magellanic Clouds are
biased towards quite active regions as they are based on OB
supergiants.  The systematic behavior of the curves studied in this
paper hint at the existence of a multiparameter relationship (possibly
dependent on $R_V$ and $N(HI)/A(V)$) describing both quiescent
(CCM-like) and active extinction curves.

The different biases between the majority of the measured Milky Way
and Magellanic Cloud extinction curves point to a weakness in our
understanding of dust properties.  The Magellanic Cloud extinction
curves are biased towards active star forming regions (OB supergiant
sightlines) and low dust columns.  The Milky Way sightlines are less
biased towards active star forming regions as they are mainly measured
along sightlines towards OB main sequence stars.  They are limited in
their variety of dust properties probed as the measurements are
generally limited to our local region of the Milky Way.  Evidence that
the type of dust seen in the Magellanic Clouds does exist in the Milky
Way is seen along Milky Way low density sightlines \citep{cla00}.
Higher dust column sightlines and sightlines towards main sequence
stars need to be measured in the Magellanic Clouds to allow us to
truly explore the possible range of dust properties.

The results of this work and those of \citet{cla00} imply that the
common usage of discussing Milky Way, LMC, and SMC dust as separate is
inaccurate.  In reality, a continuum of dust properties exists between
that seen in the quiescent and active regions (eg., see
Fig.~\ref{fig_bump_c2}).  The usual average curves can then be
arranged into a rough version of this sequence and it would be Milky
Way (CCM) -- LMC Average -- SMC Wing (AzV 456) -- LMC2 Supershell --
SMC Bar.  Ordering all the individual known extinction curves would
produce an even greater mixing of Milky Way, LMC, and SMC curves.
This work describes a qualitative view of the continuum of dust
extinction curves.  Quantifying the variation between quiescent and
active region extinction curves would give valuable clues to the
identities of dust grain properties.

\acknowledgements Support for proposal \#8198 was provided by NASA
through a grant from the Space Telescope Science Institute, which is
operated by the Association of Universities for Research in Astronomy,
Inc., under NASA contract NAS 5-26555.  We thank the referee for
comments which motivated us to improve the presentation of this paper.
This work was partially supported through JPL contract \#960785.  AUL
acknowledges support from NSF grant AST0097895.  This publication
makes use of data products from the Two Micron All Sky Survey, which
is a joint project of the University of Massachusetts and the Infrared
Processing and Analysis Center/California Institute of Technology,
funded by the National Aeronautics and Space Administration and the
National Science Foundation.  This research made use of the SIMBAD
database and VizeiR catalogue access tool, CDS, Strasbourg, France.

\appendix

\section{$R_V$ value and uncertainty equations}

The derivation of the equations to determine value of $R_V$ and its
uncertainty are given in this appendix.  To the knowledge of the
authors, these equations have not appeared in the literature before
and are included in this paper as they may be of use to others.  The
equations below are equivalent to determining $R_V$ from individual
colors and averaging the results weighted by the appropriate
uncertainties.

The equation giving the $\chi^2$ between the measured extinction curve
and the \citet{rie85} curve is
\begin{equation}
\chi^2 = \sum_i \left[ \frac{y(\lambda_i) -
   y(R_V,\lambda_i)}{\sigma(\lambda_i)} \right]^2
\end{equation}
where
\begin{eqnarray}
y(\lambda_i) & = & E(\lambda_i - V)/E(B - V) \\
\sigma(\lambda_i) & = & \sigma E(\lambda_i - V)/E(B - V) \\
y(R_V,\lambda_i) & = & \left[ A(\lambda_i)/A(V) - 1 \right] R_V
\end{eqnarray}
and $A(\lambda_i)/A(V)$ is given in table 3 of \citet{rie85}.
Differentiating $\chi^2$ with respect to $R_V$ and setting the result
to zero gives 
\begin{equation}
R_V = \frac{\sum_i \left[ A(\lambda_i)/A(V) - 1 \right]}
           {\sum_i \left[ y(\lambda_i)/\sigma(\lambda_i)^2 \right]}.
\end{equation}
Using eq.~6.19 of \citet{bev92} gives the uncertainty in $R_V$ as
\begin{equation}
\sigma(R_V)^2 = \frac{\sum_i \left[ A(\lambda_i)/A(V) - 1 \right]}
                {\sum_i \left\{ \left[ A(\lambda_i)/A(V) - 1 \right]^2 / 
                   \sigma(\lambda_i)^2 \right\} }.
\end{equation}

\begin{deluxetable}{lrrrrrrrr}
\tablewidth{0pt}
\rotate
\tabletypesize{\scriptsize}
\tablecaption{Stellar Photometry \label{tab_star_data}}
\tablehead{\colhead{star} & \colhead{V} & \colhead{B-V} & 
   \colhead{U-B} & \colhead{V-R} & \colhead{V-I} &
   \colhead{J} & \colhead{H} & \colhead{K$_s$} }
\startdata
\multicolumn{9}{c}{SMC} \\ \tableline
AzV 18 & $12.420 \pm 0.044$ & $0.041 \pm 0.006$ & $-0.794 \pm 0.021$ & $0.069 \pm 0.005$ & $0.123 \pm 0.008$ & $12.368 \pm 0.032$ & $12.336 \pm 0.030$ & $12.261 \pm 0.033$ \\
AzV 23 & $12.244 \pm 0.004$ & $0.084 \pm 0.002$ & $-0.672 \pm 0.008$ & $0.092 \pm 0.003$ & $0.188 \pm 0.006$ & $12.011 \pm 0.032$ & $11.923 \pm 0.025$ & $11.913 \pm 0.035$ \\
AzV 70 & $12.413 \pm 0.013$ & $-0.154 \pm 0.013$ & $-1.003 \pm 0.016$ & $-0.046 \pm 0.011$ & $-0.124 \pm 0.017$ & $12.711 \pm 0.032$ & $12.765 \pm 0.033$ & $12.832 \pm 0.039$ \\
AzV 214 & $13.416 \pm 0.013$ & $0.038 \pm 0.007$ & $-0.803 \pm 0.007$ & $0.065 \pm 0.004$ & $0.129 \pm 0.007$ & $13.357 \pm 0.035$ & $13.374 \pm 0.039$ & $13.312 \pm 0.048$ \\
AzV 289 & $12.396 \pm 0.026$ & $-0.118 \pm 0.009$ & $-0.984 \pm 0.013$ & $-0.032 \pm 0.005$ & $-0.111 \pm 0.014$ & $12.657 \pm 0.024$ & $12.670 \pm 0.038$ & $12.718 \pm 0.035$ \\
AzV 380 & $13.534 \pm 0.007$ & $-0.109 \pm 0.010$ & $-0.918 \pm 0.009$ & $-0.013 \pm 0.008$ & $-0.037 \pm 0.009$ & $13.747 \pm 0.031$ & $13.781 \pm 0.045$ & $13.841 \pm 0.057$ \\
AzV 398 & $13.889 \pm 0.026$ & $0.100 \pm 0.022$ & $-0.820 \pm 0.021$ & $0.107 \pm 0.007$ & $0.150 \pm 0.032$ & $13.687 \pm 0.029$ & $13.590 \pm 0.038$ & $13.374 \pm 0.042$ \\
AzV 404 & $12.197 \pm 0.009$ & $-0.098 \pm 0.006$ & $-0.825 \pm 0.005$ & $0.000 \pm 0.004$ & $-0.017 \pm 0.003$ & $12.394 \pm 0.033$ & $12.438 \pm 0.038$ & $12.409 \pm 0.037$ \\
AzV 456 & $12.888 \pm 0.019$ & $0.109 \pm 0.009$ & $-0.785 \pm 0.015$ & $0.085 \pm 0.002$ & $0.162 \pm 0.006$ & $12.820 \pm 0.020$ & $12.800 \pm 0.025$ & $12.800 \pm 0.019$ \\
AzV 462 & $12.566 \pm 0.017$ & $-0.126 \pm 0.012$ & $-0.914 \pm 0.014$ & $-0.039 \pm 0.004$ & $-0.124 \pm 0.005$ & $12.890 \pm 0.020$ & $12.940 \pm 0.025$ & $12.950 \pm 0.019$ \\ \tableline
\multicolumn{9}{c}{LMC} \\ \tableline
Sk -65 15 & $12.140 \pm 0.020$ & $-0.100 \pm 0.020$ & $-0.920 \pm 0.040$ & \nodata & \nodata & $12.417 \pm 0.025$ & $12.409 \pm 0.026$ & $12.423 \pm 0.030$ \\
Sk -65 63 & $12.560 \pm 0.020$ & $-0.160 \pm 0.020$ & $-1.020 \pm 0.040$ & \nodata & \nodata & $12.933 \pm 0.033$ & $12.941 \pm 0.036$ & $13.024 \pm 0.045$ \\
Sk -66 19 & $12.790 \pm 0.020$ & $0.120 \pm 0.020$ & $-0.780 \pm 0.040$ & \nodata & \nodata & $12.496 \pm 0.031$ & $12.382 \pm 0.028$ & $12.359 \pm 0.040$ \\
Sk -66 35 & $11.550 \pm 0.020$ & $-0.070 \pm 0.020$ & $-0.880 \pm 0.040$ & \nodata & \nodata & $11.732 \pm 0.027$ & $11.725 \pm 0.024$ & $11.700 \pm 0.034$ \\
Sk -66 88 & $12.700 \pm 0.020$ & $0.200 \pm 0.020$ & $-0.650 \pm 0.040$ & \nodata & \nodata & $12.170 \pm 0.027$ & $12.089 \pm 0.030$ & $11.945 \pm 0.033$ \\
Sk -66 106 & $11.720 \pm 0.020$ & $-0.080 \pm 0.020$ & $-0.910 \pm 0.040$ & \nodata & \nodata & $11.919 \pm 0.028$ & $11.931 \pm 0.033$ & $11.918 \pm 0.031$ \\
Sk -66 118 & $11.810 \pm 0.020$ & $-0.050 \pm 0.020$ & $-0.860 \pm 0.040$ & \nodata & \nodata & $12.072 \pm 0.026$ & $11.984 \pm 0.035$ & $12.013 \pm 0.036$ \\
Sk -66 169 & $11.560 \pm 0.020$ & $-0.130 \pm 0.020$ & $-1.000 \pm 0.040$ & \nodata & \nodata & $11.862 \pm 0.033$ & $11.895 \pm 0.038$ & $11.882 \pm 0.040$ \\
Sk -67 2 & $11.260 \pm 0.020$ & $0.080 \pm 0.020$ & $-0.690 \pm 0.040$ & \nodata & \nodata & $11.054 \pm 0.024$ & $11.010 \pm 0.034$ & $10.895 \pm 0.030$ \\
Sk -67 5 & $11.340 \pm 0.020$ & $-0.120 \pm 0.020$ & $-0.950 \pm 0.040$ & \nodata & \nodata & $11.621 \pm 0.028$ & $11.623 \pm 0.031$ & $11.636 \pm 0.032$ \\
Sk -67 36 & $12.029 \pm 0.003$ & $-0.090 \pm 0.007$ & $-0.853 \pm 0.008$ & $-0.003 \pm 0.006$ & $-0.013 \pm 0.011$ & $12.095 \pm 0.031$ & $12.116 \pm 0.026$ & $12.174 \pm 0.035$ \\
Sk -67 78 & $11.260 \pm 0.020$ & $-0.040 \pm 0.020$ & $-0.730 \pm 0.040$ & \nodata & \nodata & $11.439 \pm 0.058$ & \nodata & $11.258 \pm 0.031$ \\
Sk -67 100 & $11.950 \pm 0.020$ & $-0.090 \pm 0.020$ & $-0.860 \pm 0.040$ & \nodata & \nodata & $12.158 \pm 0.031$ & $12.235 \pm 0.031$ & $12.204 \pm 0.038$ \\
Sk -67 168 & $12.080 \pm 0.020$ & $-0.170 \pm 0.020$ & $-1.000 \pm 0.040$ & \nodata & \nodata & $12.491 \pm 0.029$ & $12.461 \pm 0.033$ & $12.528 \pm 0.036$ \\
Sk -67 228 & $11.490 \pm 0.020$ & $-0.050 \pm 0.020$ & $-0.820 \pm 0.040$ & \nodata & \nodata & $11.574 \pm 0.033$ & $11.576 \pm 0.035$ & $11.529 \pm 0.037$ \\
Sk -67 256 & $11.900 \pm 0.020$ & $-0.080 \pm 0.020$ & $-0.890 \pm 0.040$ & \nodata & \nodata & $11.938 \pm 0.032$ & $12.047 \pm 0.039$ & $11.955 \pm 0.031$ \\
Sk -68 23 & $12.810 \pm 0.020$ & $0.220 \pm 0.020$ & $-0.610 \pm 0.040$ & \nodata & \nodata & $12.180 \pm 0.027$ & $12.047 \pm 0.028$ & $11.958 \pm 0.034$ \\
Sk -68 26 & $11.630 \pm 0.003$ & $0.116 \pm 0.002$ & $-0.776 \pm 0.001$ & $0.115 \pm 0.003$ & $0.238 \pm 0.007$ & $11.410 \pm 0.060$ & $11.280 \pm 0.040$ & $11.150 \pm 0.050$ \\
Sk -68 40 & $11.710 \pm 0.020$ & $-0.070 \pm 0.020$ & $-0.790 \pm 0.040$ & \nodata & \nodata & $11.706 \pm 0.027$ & $11.716 \pm 0.037$ & $11.760 \pm 0.030$ \\
Sk -68 41 & $12.000 \pm 0.020$ & $-0.140 \pm 0.020$ & $-0.960 \pm 0.040$ & \nodata & \nodata & $12.204 \pm 0.027$ & $12.284 \pm 0.036$ & $12.242 \pm 0.032$ \\
Sk -68 129 & $12.770 \pm 0.020$ & $0.030 \pm 0.020$ & $-0.840 \pm 0.040$ & \nodata & \nodata & $12.566 \pm 0.032$ & $12.572 \pm 0.034$ & $12.514 \pm 0.041$ \\
Sk -68 140 & $12.720 \pm 0.020$ & $0.060 \pm 0.020$ & $-0.830 \pm 0.040$ & \nodata & \nodata & $12.479 \pm 0.018$ & $12.432 \pm 0.018$ & $12.380 \pm 0.020$ \\
Sk -68 155 & $12.720 \pm 0.020$ & $0.030 \pm 0.020$ & $-0.820 \pm 0.040$ & \nodata & \nodata & $12.723 \pm 0.033$ & $12.630 \pm 0.039$ & $12.669 \pm 0.035$ \\
Sk -69 108 & $12.100 \pm 0.020$ & $0.270 \pm 0.020$ & $-0.490 \pm 0.040$ & \nodata & \nodata & $11.530 \pm 0.029$ & \nodata & $11.263 \pm 0.015$ \\
Sk -69 206 & $12.840 \pm 0.020$ & $0.140 \pm 0.020$ & $-0.760 \pm 0.040$ & \nodata & \nodata & $12.408 \pm 0.033$ & $12.382 \pm 0.037$ & $12.257 \pm 0.043$ \\
Sk -69 210 & $12.590 \pm 0.020$ & $0.360 \pm 0.020$ & $-0.590 \pm 0.040$ & \nodata & \nodata & $11.796 \pm 0.036$ & $11.711 \pm 0.033$ & $11.587 \pm 0.038$ \\
Sk -69 213 & $11.970 \pm 0.020$ & $0.100 \pm 0.020$ & $-0.750 \pm 0.040$ & \nodata & \nodata & $11.698 \pm 0.033$ & $11.713 \pm 0.035$ & $11.642 \pm 0.037$ \\
Sk -69 228 & $12.120 \pm 0.020$ & $0.050 \pm 0.020$ & $-0.760 \pm 0.040$ & \nodata & \nodata & $12.029 \pm 0.034$ & $11.998 \pm 0.034$ & $11.937 \pm 0.037$ \\
Sk -69 256 & $12.610 \pm 0.020$ & $0.030 \pm 0.020$ & $-0.830 \pm 0.040$ & \nodata & \nodata & $12.799 \pm 0.032$ & $12.728 \pm 0.047$ & $12.732 \pm 0.044$ \\
Sk -69 265 & $11.880 \pm 0.020$ & $0.120 \pm 0.020$ & $-0.630 \pm 0.040$ & \nodata & \nodata & $11.680 \pm 0.029$ & $11.628 \pm 0.045$ & $11.598 \pm 0.037$ \\
Sk -69 270 & $11.270 \pm 0.020$ & $0.140 \pm 0.020$ & $-0.520 \pm 0.040$ & \nodata & \nodata & $11.009 \pm 0.036$ & $10.989 \pm 0.037$ & $10.936 \pm 0.036$ \\
Sk -69 279 & $12.790 \pm 0.020$ & $0.050 \pm 0.020$ & $-0.840 \pm 0.040$ & \nodata & \nodata & $12.701 \pm 0.036$ & $12.613 \pm 0.051$ & $12.550 \pm 0.043$ \\
Sk -69 280 & $12.660 \pm 0.020$ & $0.090 \pm 0.020$ & $-0.740 \pm 0.040$ & \nodata & \nodata & $12.492 \pm 0.031$ & $12.451 \pm 0.037$ & $12.410 \pm 0.037$ \\
Sk -70 116 & $12.050 \pm 0.020$ & $0.110 \pm 0.020$ & $-0.720 \pm 0.040$ & \nodata & \nodata & $11.679 \pm 0.034$ & $11.585 \pm 0.035$ & $11.520 \pm 0.031$ \\
Sk -70 120 & $11.590 \pm 0.020$ & $-0.060 \pm 0.020$ & $-0.880 \pm 0.040$ & \nodata & \nodata & $11.831 \pm 0.033$ & $11.803 \pm 0.028$ & $11.835 \pm 0.036$ \\
\enddata
\end{deluxetable}

\begin{deluxetable}{llcccr}
\tablewidth{0pt}
\tabletypesize{\footnotesize}
\tablecaption{Extinction Curve Data \label{tab_ext_data}}
\tablehead{\colhead{reddened} & \colhead{comparison} & 
   \colhead{$E(B-V)$} & \colhead{$R_V$} & \colhead{N(\ion{H}{1})} & 
   \colhead{N(\ion{H}{1})/A(V)} \\
   \colhead{star} & \colhead{star} & [mag] & & 
   [$1\times 10^{21}$] & [$1\times 10^{21}$] }
\startdata
\multicolumn{6}{c}{SMC - Bar sample} \\ \tableline
AzV 18 & AzV 462 & $0.167 \pm 0.013$ & $3.30 \pm 0.38$ & $8.50 \pm 0.50$ & $15.41 \pm 2.33$ \\
AzV 23 & AzV 404 & $0.182 \pm 0.006$ & $2.65 \pm 0.17$ & $8.50 \pm 0.50$ & $17.65 \pm 1.64$ \\
AzV 214 & AzV 380 & $0.147 \pm 0.012$ & $2.40 \pm 0.30$ & $2.50 \pm 0.50$ & $7.08 \pm 1.77$ \\
AzV 398 & AzV 289 & $0.218 \pm 0.024$ & $3.14 \pm 0.34$ & $8.00 \pm 2.00$ & $11.69 \pm 3.43$ \\
\multicolumn{2}{l}{average} & \nodata & $2.74 \pm 0.13$  & \nodata & $13.18 \pm 1.02$ \\  \tableline
\multicolumn{6}{c}{SMC - Wing sample} \\ \tableline
AzV 456 & AzV 70 & $0.263 \pm 0.016$ & $2.05 \pm 0.17$ & $4.00 \pm 0.50$ & $7.40 \pm 1.20$ \\ \tableline
\multicolumn{6}{c}{LMC - LMC2 supershell sample} \\ \tableline
Sk -68 140 & Sk -68 41 & $0.200 \pm 0.028$ & $3.27 \pm 0.24$ & $4.00 \pm 1.00$ & $6.12 \pm 1.81$ \\
Sk -68 155 & Sk -67 168 & $0.200 \pm 0.028$ & $2.83 \pm 0.23$ & $5.00 \pm 1.00$ & $8.82 \pm 2.28$ \\
Sk -69 228 & Sk -65 15 & $0.150 \pm 0.028$ & $3.35 \pm 0.33$ & $3.50 \pm 0.50$ & $6.97 \pm 1.79$ \\
Sk -69 256 & Sk -68 41 & $0.170 \pm 0.028$ & $0.64 \pm 0.19$ & $2.50 \pm 0.50$ & $23.10 \pm 9.24$ \\
Sk -69 265 & Sk -68 40 & $0.190 \pm 0.028$ & $1.68 \pm 0.19$ & $5.00 \pm 0.50$ & $15.71 \pm 3.35$ \\
Sk -69 270 & Sk -67 228 & $0.190 \pm 0.028$ & $2.34 \pm 0.22$ & $3.50 \pm 1.00$ & $7.86 \pm 2.64$ \\
Sk -69 279 & Sk -65 63 & $0.210 \pm 0.028$ & $3.33 \pm 0.26$ & $4.00 \pm 1.00$ & $5.73 \pm 1.69$ \\
Sk -69 280 & Sk -67 100 & $0.180 \pm 0.028$ & $3.12 \pm 0.27$ & $6.00 \pm 1.00$ & $10.68 \pm 2.62$ \\
Sk -70 116 & Sk -67 256 & $0.190 \pm 0.028$ & $3.41 \pm 0.27$ & $3.50 \pm 0.50$ & $5.39 \pm 1.19$ \\
\multicolumn{2}{l}{average\tablenotemark{a}} & \nodata & $2.76 \pm 0.09$  & \nodata & $6.97 \pm 0.67$ \\  \tableline
\multicolumn{6}{c}{LMC - Average sample} \\ \tableline
Sk -66 19 & Sk -66 169 & $0.250 \pm 0.028$ & $3.44 \pm 0.21$ & $7.00 \pm 1.00$ & $8.15 \pm 1.57$ \\
Sk -66 88 & Sk -66 106 & $0.280 \pm 0.028$ & $3.67 \pm 0.19$ & $5.50 \pm 0.50$ & $5.35 \pm 0.78$ \\
Sk -67 2 & Sk -66 35 & $0.150 \pm 0.028$ & $3.62 \pm 0.35$ & $1.00 \pm 0.50$ & $1.84 \pm 1.00$ \\
Sk -68 23 & Sk -67 36 & $0.310 \pm 0.021$ & $3.35 \pm 0.13$ & $1.50 \pm 0.50$ & $1.45 \pm 0.50$ \\
Sk -68 26 & Sk -66 35 & $0.186 \pm 0.020$ & $3.43 \pm 0.24$ & $3.50 \pm 0.50$ & $5.48 \pm 1.05$ \\
Sk -68 129 & Sk -68 41 & $0.170 \pm 0.028$ & $3.36 \pm 0.30$ & $4.00 \pm 1.00$ & $7.01 \pm 2.20$ \\
Sk -69 108 & Sk -67 78 & $0.310 \pm 0.028$ & $3.15 \pm 0.16$ & $3.00 \pm 0.50$ & $3.07 \pm 0.61$ \\
Sk -69 206 & Sk -67 5 & $0.260 \pm 0.028$ & $3.68 \pm 0.21$ & $8.00 \pm 2.00$ & $8.36 \pm 2.33$ \\
Sk -69 210 & Sk -66 118 & $0.410 \pm 0.028$ & $3.32 \pm 0.12$ & $10.00 \pm 2.50$ & $7.35 \pm 1.93$ \\
Sk -69 213 & Sk -70 120 & $0.160 \pm 0.028$ & $3.96 \pm 0.36$ & $1.50 \pm 0.50$ & $2.37 \pm 0.92$ \\
\multicolumn{2}{l}{average} & \nodata & $3.41 \pm 0.06$  & \nodata & $3.25 \pm 0.28$ \\
\enddata
\tablenotetext{a}{Excluding Sk -69 256}
\end{deluxetable}

\begin{deluxetable}{lrrrrrr}
\tablewidth{0pt}
\tabletypesize{\scriptsize}
\tablecaption{FM Parameters \label{tab_fm_param}}
\tablehead{\colhead{reddened} & \multicolumn{6}{c}{FM parameters} \\
   \colhead{star} & \colhead{$c_1$} & \colhead{$c_2$} &
   \colhead{$c_3$} & \colhead{$c_4$} & \colhead{$x_o$} & 
   \colhead{$\gamma$} }
\startdata
\multicolumn{7}{c}{SMC - Bar sample} \\ \tableline
AzV 18 & $-4.938 \mp 0.634$ & $2.267 \pm 0.204$ & $0.362 \pm 0.190$ & $0.176 \pm 0.084$ & $4.600 \pm 0.000$ & $1.000 \pm 0.000$ \\
AzV 23 & $-5.170 \mp 0.289$ & $2.382 \pm 0.092$ & $0.489 \pm 0.125$ & $0.462 \pm 0.057$ & $4.600 \pm 0.000$ & $1.000 \pm 0.000$ \\
AzV 214 & $-4.495 \mp 0.687$ & $2.264 \pm 0.222$ & $-0.123 \pm 0.065$ & $-0.435 \pm 0.197$ & $4.600 \pm 0.000$ & $1.000 \pm 0.000$ \\
AzV 398 & $-5.382 \mp 0.808$ & $2.328 \pm 0.276$ & $0.314 \pm 0.167$ & $0.263 \pm 0.128$ & $4.600 \pm 0.000$ & $1.000 \pm 0.000$ \\
average & $-4.959 \mp 0.197$ & $2.264 \pm 0.040$ & $0.389 \pm 0.110$ & $0.461 \pm 0.079$ & $4.600 \pm 0.000$ & $1.000 \pm 0.000$ \\ \tableline
\multicolumn{7}{c}{SMC - Wing sample} \\ \tableline
AzV 456 & $-0.856 \mp 0.246$ & $1.038 \pm 0.074$ & $3.215 \pm 0.439$ & $0.107 \pm 0.038$ & $4.703 \pm 0.018$ & $1.212 \pm 0.019$ \\ \tableline
\multicolumn{7}{c}{LMC - LMC2 supershell sample} \\ \tableline
Sk -68 140 & $-1.547 \mp 0.509$ & $1.247 \pm 0.197$ & $1.151 \pm 0.303$ & $0.357 \pm 0.148$ & $4.464 \pm 0.039$ & $0.855 \pm 0.014$ \\
Sk -68 155 & $-2.689 \mp 0.487$ & $1.580 \pm 0.230$ & $0.923 \pm 0.170$ & $0.117 \pm 0.049$ & $4.617 \pm 0.021$ & $0.693 \pm 0.012$ \\
Sk -69 228 & $-2.443 \mp 0.662$ & $1.373 \pm 0.275$ & $0.716 \pm 0.234$ & $-0.115 \pm 0.065$ & $4.721 \pm 0.049$ & $0.643 \pm 0.011$ \\
Sk -69 256 & $-1.139 \mp 0.493$ & $1.101 \pm 0.205$ & $0.746 \pm 0.271$ & $0.257 \pm 0.133$ & $4.817 \pm 0.062$ & $0.751 \pm 0.013$ \\
Sk -69 265 & $-3.083 \mp 0.612$ & $1.509 \pm 0.237$ & $0.360 \pm 0.101$ & $-0.313 \pm 0.152$ & $4.363 \pm 0.036$ & $0.536 \pm 0.009$ \\
Sk -69 270 & $-3.926 \mp 0.755$ & $1.821 \pm 0.281$ & $3.725 \pm 0.837$ & $0.208 \pm 0.091$ & $4.509 \pm 0.034$ & $1.290 \pm 0.021$ \\
Sk -69 279 & $-2.669 \mp 0.551$ & $1.350 \pm 0.196$ & $0.978 \pm 0.248$ & $0.017 \pm 0.009$ & $4.602 \pm 0.032$ & $0.708 \pm 0.012$ \\
Sk -69 280 & $0.468 \mp 0.282$ & $0.809 \pm 0.142$ & $1.209 \pm 0.306$ & $0.232 \pm 0.130$ & $4.470 \pm 0.032$ & $0.667 \pm 0.011$ \\
Sk -70 116 & $-1.707 \mp 0.567$ & $1.153 \pm 0.192$ & $6.557 \pm 1.504$ & $0.193 \pm 0.080$ & $4.387 \pm 0.047$ & $1.878 \pm 0.031$ \\
average\tablenotemark{a} & $-1.475 \mp 0.152$ & $1.132 \pm 0.029$ & $1.463 \pm 0.121$ & $0.294 \pm 0.057$ & $4.558 \pm 0.021$ & $0.945 \pm 0.026$ \\ \tableline
\multicolumn{7}{c}{LMC - Average sample} \\ \tableline
Sk -66 19 & $-0.724 \mp 0.282$ & $0.902 \pm 0.116$ & $3.036 \pm 0.465$ & $0.276 \pm 0.118$ & $4.567 \pm 0.027$ & $1.132 \pm 0.030$ \\
Sk -66 88 & $-0.960 \mp 0.276$ & $1.051 \pm 0.115$ & $2.095 \pm 0.310$ & $0.113 \pm 0.042$ & $4.580 \pm 0.016$ & $0.916 \pm 0.015$ \\
Sk -67 2 & $-3.914 \mp 0.851$ & $1.781 \pm 0.347$ & $3.680 \pm 0.744$ & $0.884 \pm 0.239$ & $4.566 \pm 0.014$ & $0.996 \pm 0.016$ \\
Sk -68 23 & $-0.152 \mp 0.070$ & $0.696 \pm 0.053$ & $4.647 \pm 0.572$ & $0.453 \pm 0.108$ & $4.493 \pm 0.014$ & $1.135 \pm 0.019$ \\
Sk -68 26 & $-1.076 \mp 0.343$ & $1.082 \pm 0.130$ & $4.085 \pm 0.621$ & $0.121 \pm 0.045$ & $4.622 \pm 0.015$ & $1.076 \pm 0.018$ \\
Sk -68 129 & $-2.318 \mp 0.507$ & $1.388 \pm 0.243$ & $1.632 \pm 0.340$ & $0.468 \pm 0.162$ & $4.569 \pm 0.019$ & $0.790 \pm 0.013$ \\
Sk -69 108 & $-1.262 \mp 0.327$ & $0.992 \pm 0.102$ & $5.046 \pm 0.621$ & $0.384 \pm 0.101$ & $4.519 \pm 0.010$ & $1.079 \pm 0.017$ \\
Sk -69 206 & $-1.243 \mp 0.342$ & $1.217 \pm 0.144$ & $1.169 \pm 0.204$ & $0.016 \pm 0.007$ & $4.500 \pm 0.020$ & $0.689 \pm 0.011$ \\
Sk -69 210 & $-1.559 \mp 0.187$ & $1.182 \pm 0.087$ & $1.583 \pm 0.166$ & $0.307 \pm 0.074$ & $4.641 \pm 0.007$ & $0.720 \pm 0.011$ \\
Sk -69 213 & $-2.791 \mp 0.703$ & $1.594 \pm 0.300$ & $1.816 \pm 0.337$ & $0.527 \pm 0.177$ & $4.564 \pm 0.018$ & $0.735 \pm 0.014$ \\
average & $-0.890 \mp 0.142$ & $0.998 \pm 0.027$ & $2.719 \pm 0.137$ & $0.400 \pm 0.036$ & $4.579 \pm 0.007$ & $0.934 \pm 0.016$ \\
\enddata
\tablenotetext{a}{Excluding Sk -69 256}
\end{deluxetable}

\begin{deluxetable}{ccccc}
\tablewidth{0pt}
\tabletypesize{\footnotesize}
\tablecaption{Sample Average Curves \label{tab_ave_curves}}
\tablehead{ & & \colhead{SMC} & \colhead{LMC2} & \colhead{LMC} \\
  \colhead{$\lambda$} & \colhead{$x$} &
  \colhead{Bar} & \colhead{Supershell} & \colhead{Average}
  \\
  \colhead{[$\micron$]} & \colhead{[$\micron^{-1}$]} &
  \multicolumn{3}{c}{[$A(\lambda)/A(V)$]} }
\startdata
2.198 & 0.455 & $0.016 \pm 0.003$ & $0.101 \pm 0.003$ & $0.030 \pm 0.003$ \\
1.650 & 0.606 & $0.169 \pm 0.020$ & $0.097 \pm 0.020$ & $0.186 \pm 0.020$ \\
1.250 & 0.800 & $0.131 \pm 0.013$ & $0.299 \pm 0.013$ & $0.257 \pm 0.013$ \\
0.810 & 1.235 & $0.567 \pm 0.048$ & \nodata & \nodata \\
0.650 & 1.538 & $0.801 \pm 0.113$ & \nodata & \nodata \\
0.550 & 1.818 & $1.000 \pm 0.046$ & $1.000 \pm 0.048$ & $1.000 \pm 0.048$ \\
0.440 & 2.273 & $1.374 \pm 0.127$ & $1.349 \pm 0.113$ & $1.293 \pm 0.113$ \\
0.370 & 2.703 & $1.672 \pm 0.123$ & $1.665 \pm 0.046$ & $1.518 \pm 0.046$ \\
0.296 & 3.375 & $2.000 \pm 0.095$ & $1.899 \pm 0.127$ & $1.786 \pm 0.127$ \\
0.276 & 3.625 & $2.220 \pm 0.093$ & $2.067 \pm 0.123$ & $1.969 \pm 0.123$ \\
0.258 & 3.875 & $2.428 \pm 0.093$ & $2.249 \pm 0.095$ & $2.149 \pm 0.095$ \\
0.242 & 4.125 & $2.661 \pm 0.095$ & $2.447 \pm 0.093$ & $2.391 \pm 0.093$ \\
0.229 & 4.375 & $2.947 \pm 0.099$ & $2.777 \pm 0.093$ & $2.771 \pm 0.093$ \\
0.216 & 4.625 & $3.161 \pm 0.102$ & $2.922 \pm 0.095$ & $2.967 \pm 0.095$ \\
0.205 & 4.875 & $3.293 \pm 0.104$ & $2.921 \pm 0.099$ & $2.846 \pm 0.099$ \\
0.195 & 5.125 & $3.489 \pm 0.105$ & $2.812 \pm 0.102$ & $2.646 \pm 0.102$ \\
0.186 & 5.375 & $3.637 \pm 0.107$ & $2.805 \pm 0.104$ & $2.565 \pm 0.104$ \\
0.178 & 5.625 & $3.866 \pm 0.112$ & $2.863 \pm 0.105$ & $2.566 \pm 0.105$ \\
0.170 & 5.875 & $4.013 \pm 0.115$ & $2.932 \pm 0.107$ & $2.598 \pm 0.107$ \\
0.163 & 6.125 & $4.243 \pm 0.119$ & $3.060 \pm 0.112$ & $2.607 \pm 0.112$ \\
0.157 & 6.375 & $4.472 \pm 0.124$ & $3.110 \pm 0.115$ & $2.668 \pm 0.115$ \\
0.151 & 6.625 & $4.776 \pm 0.131$ & $3.299 \pm 0.119$ & $2.787 \pm 0.119$ \\
0.145 & 6.875 & $5.000 \pm 0.135$ & $3.408 \pm 0.124$ & $2.874 \pm 0.124$ \\
0.140 & 7.125 & $5.272 \pm 0.142$ & $3.515 \pm 0.131$ & $2.983 \pm 0.131$ \\
0.136 & 7.375 & $5.575 \pm 0.148$ & $3.670 \pm 0.135$ & $3.118 \pm 0.135$ \\
0.131 & 7.625 & $5.795 \pm 0.153$ & $3.862 \pm 0.142$ & $3.231 \pm 0.142$ \\
0.127 & 7.875 & $6.074 \pm 0.160$ & $3.937 \pm 0.148$ & $3.374 \pm 0.148$ \\
0.123 & 8.125 & $6.297 \pm 0.368$ & $4.055 \pm 0.153$ & $3.366 \pm 0.153$ \\
0.119 & 8.375 & $6.436 \pm 0.271$ & $3.969 \pm 0.160$ & $3.467 \pm 0.160$ \\
0.116 & 8.625 & $6.992 \pm 0.201$ & \nodata & \nodata \\
\enddata
\end{deluxetable}

\end{document}